\documentclass[aps,twocolumn,superscriptaddress,floatfix,showpacs]{revtex4}
\usepackage{amsmath,euscript,theorem,graphicx,rotating}
\usepackage[letterpaper,hmargin=25mm,vmargin=25mm]{geometry}
\usepackage[latin1]{inputenc}
\usepackage[T1]{fontenc}
\usepackage{txfonts}
\usepackage{url}
\usepackage{xr} % for eXternal References
\usepackage{color}
\usepackage{dcolumn}% Align table columns on decimal point
\usepackage{bm}% bold math
\usepackage{longtable}% To allow splitting tables across multiple pages
\usepackage{xspace}% To preserve spaces after macros
\usepackage{booktabs}
\usepackage{version}
\usepackage{natbib,bibentry}

\usepackage{array}
\newcolumntype{L}{>{$}l<{$}}
\newcolumntype{R}{>{$}r<{$}}
\newcolumntype{C}{>{$}c<{$}}

\newcommand{\rr}{{\mathbf r}}

\newcommand{\RR}{{\mathbf R}}
\newcommand{\rp}{{\mathbf r'}}

\newcommand{\etal}{\emph{et al.}\xspace}

\newcommand{\order}[1]{\ensuremath{\mathcal{O}(#1)}\xspace}
\newcommand{\Eelst}[0]{\ensuremath{E^{(1)}_{\rm elst}}\xspace}
\newcommand{\EelstDM}[0]{\ensuremath{E^{(1)}_{\rm elst}[{\rm DM}]}\xspace}
\newcommand{\Eexch}[0]{\ensuremath{E^{(1)}_{\rm exch}}\xspace}

\newcommand{\CamCASP}{{\sc CamCASP}\xspace}
\newcommand{\ORIENT}{{\sc Orient}\xspace}
\newcommand{\SAPT}{{\sc Sapt2008}\xspace}

\newcommand{\DALTON}{{\sc DALTON} 2.0\xspace}

\newcommand{\w}[1]{\ensuremath{w^{#1}}\xspace}
\newcommand{\R}[1]{\ensuremath{\rho^{#1}}\xspace}
\newcommand{\wL}[1]{\ensuremath{w_{\rm L}^{#1}}\xspace}
\newcommand{\wT}[1]{\ensuremath{\tilde{w}^{#1}}\xspace}
\newcommand{\DFISAfunc}[1]{\ensuremath{\Delta_{\text{DF-ISA}}(#1)}\xspace}
\newcommand{\wIT}[2]{\ensuremath{w^{#1}_{#2}}\xspace}
\newcommand{\ISAfunc}[1]{\ensuremath{\Delta_{\rm stock(#1)}}\xspace}

\newcommand{\ISA}[0]{BS-ISA\xspace}
\newcommand{\DFISA}[0]{BS-ISA+DF\xspace}

\newcommand{\rhot}[0]{\ensuremath{\tilde{\rho}}\xspace}

\mathchardef\lt="313C \mathchardef\gt="313E
\mathcode`<="4268 \mathcode`>="5269
\newcommand{\JCP}[0]{J. Chem. Phys.\ }

\newcommand{\JCTC}[0]{J. Chem. Theory Comput.\ }

\newcommand{\CPL}[0]{Chem. Phys. Lett.\ }

\newcommand{\PRB}[0]{Phys. Rev. B\ }

\newcommand{\MolP}[0]{Mol. Phys.\ }

\newcommand{\PCCP}[0]{Phys. Chem. Chem. Phys.\ }

\newcommand{\IRPC}[0]{Int. Revs. Phys. Chem.\ }

\newcommand{\CGD}[0]{Cryst. Growth Des.\ }

\newcommand{\ChemComm}[0]{Chem. Commun.\ }

\begin{document}

\title{Distributed multipoles from a robust basis-space implementation of the 
iterated stockholder atoms procedure}

\author{Alston J. Misquitta}
\affiliation{School of Physics and Astronomy, Queen Mary, University of London,
London E1 4NS, U.K.}
\author{Anthony J. Stone}
\affiliation{University Chemical Laboratory, Lensfield Road,
Cambridge, CB2 1EW, U.K.}
\author{Farhang Fazeli}
\affiliation{School of Physics and Astronomy, Queen Mary, University of London,
London E1 4NS, U.K.}

\date{\today}

\begin{abstract}
The recently developed iterated stockholder atoms (ISA) approach of 
Lillestolen and Wheatley (Chem. Commun. {\bf 2008}, 5909 (2008)) offers a powerful 
method for defining atoms in a molecule. However, the real-space algorithm
is known to converge very slowly, if at all.
Here we present a robust, basis-space algorithm of the ISA method 
and demonstrate its applicability on a variety of systems.
We show that this algorithm exhibits rapid convergence 
(taking around 10--80 iterations) 
with the number of iterations needed being unrelated to the system
size or basis set used.
Further, we show that the multipole moments calculated using this basis-space ISA 
method are as good as, or better than those obtained from Stone's
distributed multipole analysis (J. Chem. Theory Comput. {\bf 1}, 1128 (2005) ),
exhibiting better convergence properties and resulting in
better behaved penetration energies.
This can have significant consequences in the development of intermolecular
interaction models.
\end{abstract}

%PACS
%31.10.+z       Theory of electronic structure, electronic transitions, and
%chemical binding
%31.15.-p        Calculations and mathematical techniques in atomic and
%molecular physics
%31.15.ap       Polarizabilities and other atomic and molecular properties
%overlap, Hückel, PPP methods, etc.)
%34.20.Gj       Intermolecular and atom-molecule potentials and forces


\maketitle

\section{Introduction}
\label{sec:introduction}
The concept of atoms in a molecule (AIM) underlies much of our 
scientific understanding and almost all classical models of atomic and
molecular interactions. Almost all force fields
are built up from pair-wise atom--atom interactions. Many-body
non-additive effects, when they are included, are generally added as a 
correction to this atom--atom picture. This viewpoint has its
limitations: it is inappropriate for metals, and gets progressively more
inadequate as the electron delocalization length increases
\cite{MisquittaSSA10}. Nevertheless, it is likely that even for these systems,
atom--atom models may be useful when augmented with continuum models that account for the
metallic component.

The problem with the AIM model is that there is no unique way to define
an atom in a molecule. While some atomic properties (e.g., spectral transitions)
may be preserved upon chemical bonding, others (e.g., atomic size and charge) 
are lost or alter dramatically. Some of these are not physical
observables, and methods of defining and calculating them remain controversial.

Many techniques for identifying atoms within molecules are concerned
only with determining point 
charges that reproduce the electrostatic potential of the molecule
reasonably accurately \cite[e.g., ][]{GuerraHBB04,BultinckCN09}.
We are concerned here with obtaining a well-defined specification of
an atom within a molecule that provides an accurate and
rapidly-converging multipole expansion of the electrostatic field
around the atom. For this, we need a description of each atom that is
as nearly as possible spherical.
This condition rules out AIM approaches such as that
of Bader \cite{Bader90}, which is very well defined and has some
useful properties, but which leads to highly
non-spherical atomic shapes that result in 
multipole expansions with poor convergence properties \cite{Jansen:96,KosovP00,JoubertP02}.
Of the AIM methods that result in nearly-spherical atoms, the
Hirshfeld stockholder method is one of the most popular. In this
method, given functions $\w{a}(\rr)$ describing the
spherically-averaged electron density of the free 
atoms, we define the atom in the molecule $\R{a}(\rr)$ using: 
\begin{align}
  \R{a}(\rr) = \rho(\rr) \times \frac{\w{a}(\rr)}{\sum_b \w{b}(\rr)},
    \label{eq:stockholder}
\end{align}
where $\rho$ is the total molecular density and the indices $a$ and $b$ label
the atoms. However this method has two disadvantages: (1) it requires pre-computed
shape functions obtained from free atom calculations, and (2) as a result,
it is unable to respond to changes in the atomic densities due to 
chemical bonding. The last point is subtle and requires an explanation.
A free carbon atom is more diffuse than a free oxygen atom and this 
difference is reflected in the Hirshfeld shape functions of these atoms.
However, on formation of the C--O bond, say in carbon dioxide, the carbon
atom must be more compact and the oxygens more diffuse, as the more
electronegative oxygen atoms draw some of the electronic density away
from the carbon atom. The Hirshfeld procedure does not take account
of this phenomenon.
% \edit{AJM: I think the Hirshfeld approach would result in a 
% negative carbon atom in CO$_2$, but don't have the code to verify
% this. AJS: Reworded to avoid the need for this.}

Recently, Lillestolen and Wheatley \cite{LillestolenW08} proposed
a novel and rather appealing alternative to the conventional Hirshfeld
procedure. In their iterated stockholder atoms (ISA) method no free-atom 
shape functions $\w{a}(\rr)$ are needed. Instead, all we do is assume the
existence of these spherically symmetric shapes that are, by definition,
required to be spherical averages of the atomic densities defined in 
eqn.~\eqref{eq:stockholder}, that is,
\begin{align}
  \w{a}(\rr) = < \R{a}(\rr) >_{\rm sph},
  \label{eq:ISA-sphavg}
\end{align}
where the angle bracket indicates the spherical average. The idea here is to make
an initial guess for the shape functions \w{a}, with the only restrictions on them
being that they are integrable and positive over all space, and then to
iterate eqns.~\eqref{eq:stockholder} and \eqref{eq:ISA-sphavg} until 
the shape functions attain a desired convergence.

Using a real-space implementation of the ISA algorithm, Lillestolen and Wheatley
showed that this scheme always converges to a unique solution, and that
the atomic charges obtained from the ISA method appear to reproduce the
molecular dipole moments better than other distribution schemes. This is
promising; if it is generally true, the nearly-spherical ISA atoms could be useful
for the construction of compact distributed multipole schemes for molecular systems.
If this were the only feature of the ISA atoms, this method would be no 
different from the standard Hirshfeld technique. However, as we shall see, the ISA
atoms additionally capture changes that we commonly associate with chemical bonding.
This effect is totally absent in the usual Hirshfeld approach.

The ISA partitioning method is not limited to the density only, but, in a 
straightforward generalization, can be used to obtain distributed second-order 
quantities such as the frequency-dependent polarizabilities. However, to achieve
this, we cannot use a real-space, grid-based ISA algorithm due to the relatively
large computational cost of grid-based methods, and the prohibitively large 
number of transition densities we would have to partition. 
Consequently we need a basis-space
implementation. A further motivation for this is that, although the
Lillestolen--Wheatley method is guaranteed to converge, in practice it
is found to converge very slowly\cite{VerstraelenAvanSW12}.

% \edit{
% I just thought of a possibly unusual way of testing the DMA and ISA methods: create a
% `molecule' from un-relaxed atomic functions, say by taking two CO$_2$ molecular
% densities and placing them in close proximity. These are unrelaxed, so the combined 
% MO matrix will be block diagonal. In the absence of any relaxation,
% the moments of this system should be those of the two individual molecules
% summed up. In particular, site charges and higher order moments should be
% unchanged. Do the two methods reproduce this? I think the original 
% DMA (switch = 0) method will result in unchanged multipoles on the molecules,
% but the new one won't. It's not clear to me what the DF-ISA method will give,
% but in this case we know what the correct charges/moments should be so this 
% pseudo molecule could be a good testing ground for the methods.}

\section{Theory and numerical details}
\label{sec:theory}
The basic idea in the basis-space approach is to use expansions for all the
quantities that appear in eqns.~\eqref{eq:stockholder} and \eqref{eq:ISA-sphavg}.
Our goal is to construct an appropriate functional that allows us to
obtain the expansion coefficients. The density will normally be 
expanded in an auxiliary basis set using standard density-fitting techniques:
\begin{align}
  \rhot(\rr) = \sum_k d_k \ \chi_k(\rr).
  \label{eq:rho-expansion}
\end{align}
Here, $d_k$ are the expansion coefficients and $\chi_k$ the auxiliary basis set.
The density fitting is performed by minimizing the functional
\begin{align}
  \Delta^{\rm DF} = \iint (\rho(\rr) - \rhot(\rr))
         \frac{1}{|\rr-\rp|}(\rho(\rp) - \rhot(\rp)) \ d\rr d\rp,
         \label{eq:DF-functional}
\end{align}
where $\rho$ is the non-expanded density, which, for closed-shell systems
can be written in terms of the occupied molecular orbitals $\phi_i$ as
$\rho = 2 \sum_i |\phi_i|^2$. As we have done in Ref.~\onlinecite{MisquittaS06},
we can enforce charge conservation by including the following constraint:
\begin{align}
  \Delta^{\rm Q} = \lambda \left(
          \int \rhot(\rr) \ d\rr - N
                           \right)^2,
      \label{eq:Q-functional}
\end{align}
where $N$ is the total number of electrons in the molecule and $\lambda$ is 
the weight given to the charge-conservation functional. We typically set 
$\lambda = 1000$.

The expansion for the atomic density \R{a} is given by
\begin{align}
  \R{a}(\rr) = \sum_k c_k^a \ \xi_k^a(\rr),
  \label{eq:rhoA-expansion}
\end{align}
where the $\xi_k^a$ are basis functions associated with site $a$ (these will 
normally be gaussian-type orbitals (GTOs) centred at $a$) and the coefficients
$c_k^a$ are to be determined by minimizing an appropriate ISA functional.
The basis set used for the atomic expansion of site $a$
will normally be a subset of the auxiliary basis used in
the density fitting, limited to only include functions centered on site
$a$. However, due to the the differences in the density-fitting and
ISA functionals, we should expect to use different basis sets for each.
The main reason for this is that the density fitting is performed
using the entire monomer auxiliary basis set: basis deficiencies at a site can
be somewhat made up for using auxiliary functions from neighbouring
sites. This flexibility is not present in the ISA functional. Additionally,
as we shall see, the ISA functionals require considerable flexibility in
the AIM density tails. 

Given an atomic density \R{a} expanded in a basis of spherical GTOs, 
the atomic shape function \w{a} is trivially defined as just the 
$s$-function ($l=0$) part of \R{a}. That is, with the above expansion for \R{a}
we get
\begin{align}
  \w{a}(\rr) = \sum_{k \in \text{s-func}} c_k^a \; \xi_{k,{\rm s}}^{a}(\rr),
  \label{eq:w-expansion}
\end{align}
where we have emphasised the $s$-character of the expansion functions
with the additional `s' in the subscript. In this manner, the ISA
spherical average step, eq.~\eqref{eq:ISA-sphavg}, is trivial when using
basis expansions for the atomic functions. By contrast, it
is cumbersome to implement on a grid.

Recently Verstraelen \etal \cite{VerstraelenAvanSW12} have described
a similar approach which they call the Gaussian-ISA, or GISA, method.
The GISA method is formally exactly the same as the
Lillestolen-Wheatley ISA except that the shape functions are described
using an expansion in a series of s-functions as is proposed here.
However our proposal differs from the GISA method in important ways,
which we will detail below.

The basis-space implementation of the stockholder partitioning step in 
eq.~\eqref{eq:stockholder} is not unique and can be achieved using a
variety of functionals. The obvious choice is to minimize
\begin{align}
  \ISAfunc{A} = \sum_{a} \left\| 
         \biggl( \R{a} - \rho \frac{\w{a}}{\sum_b \w{b}} \biggr)^2
                                 \right\|,
  \label{eq:stockholder-A-functional}
\end{align}
where $\| \cdots \|$ indicates an appropriate norm, which in 
this case must be the overlap norm, as the integrand must be evaluated 
numerically on a grid.
The following alternative allows us to use either the coulomb or overlap
norm:
\begin{align}
  \ISAfunc{B} = \sum_{a} \biggl\| 
              \biggl( \R{a} \sum_b \w{b} - \rho \w{a} \biggr)^2
                                 \biggr\|.
  \label{eq:stockholder-B-functional}
\end{align}
Notice that both functionals allow us to determine the atomic densities \R{a}
one at a time. This possibility can be used to make the algorithm scale
linearly with the number of atoms.

\subsection{A na\"{i}ve ISA algorithm}
\label{ssec:naive-ISA-algorithm}

It might seem that a straightforward ISA algorithm would be:
\begin{enumerate}
  \item Initialize the shape functions \w{a}. While the starting point does not
    matter, a good starting point will ensure faster convergence. Reasonable choices
    are:
    \begin{enumerate}
      \item Set the coefficients of one GTO to be $1.0$ and the rest zero.
      \item Use the density-fitting solution to determine the starting coefficients.
    \end{enumerate}
  \item Determine all the atomic densities \R{a} using either $\ISAfunc{A}$ or 
    $\ISAfunc{B}$. This can be done one atom at a time.
  \item Update the shape functions \w{a}.
  \item Check for convergence (see below). If not converged, iterate.
\end{enumerate}
This algorithm can lead to converged shape functions, but 
when we use the standard density-fitting basis sets to describe the ISA
atomic densities, the converged $\ISAfunc{A/B}$ is non-zero, and
% we do not get $\sum_a \R{a} = \rho$ as we would expect
% from the form of the functionals \ISAfunc{A} and \ISAfunc{B}. 
the total charge is incorrect by around $0.001$ to $0.01$ electrons.
The primary cause for this discrepancy is that there is not enough flexibility
in the typical density fitting auxiliary basis sets to fit the total density 
and the ISA atomic tails simultaneously. As explained above, in a standard fit to the density
using eq.~\eqref{eq:DF-functional}, this lack of flexibility is not an issue 
as we minimize the functional $\Delta^{\text{DF}}$ in the variational space 
spanned by the molecular auxiliary basis set. 
By contrast, in the ISA procedure described above, the fit to the density
of site $a$ is performed using only basis functions at that site.
This significantly reduces the variational flexibility of
the basis, and hence results in a poor fit to the total density.

This problem is resolved by an increase in the auxiliary basis flexibility,
but this in turn leads to numerical instabilities. These are first manifested 
in small negative terms in the tail regions of the AIM densities \R{a},
which, as the n{\"a}ive ISA iterations progress, tend to grow uncontrollably
and lead to a meaningless solution. (The proof of convergence of the
ISA method\cite{LillestolenW08} fails if the weight function becomes
negative anywhere.) In the following we will describe
how the basis sets are extended, and instabilities controlled, while
simultaneously retaining the linearity of the ISA functionals.

\subsection{ISA basis sets}
\label{ssec:ISA-basis-sets}

The main problem with the auxiliary basis sets designed for density-fitting
is that their s-function block is not flexible enough to describe 
the ISA atomic density tails well enough. They need to be described very well for
the ISA solution to stabilize and converge reliably. There may also be
inadequacies in the higher angular functions, and we have some evidence
that this may be the case, but it is the s-functions that are the most 
important, due to their role in determining the AIM shape functions.
We have therefore created hybrid `DF-ISA' auxiliary basis sets that 
comprise a very flexible s-function set designed to allow
good modelling of the ISA shape-function tails, together with
the higher symmetry functions from the standard RI-MP2 density fitting
basis sets \cite{WeigendHPA98,WeigendKH02}.

There are a few requirements for a good ISA basis: it should be flexible
enough to be able to describe the ISA atomic shapes (particularly in the
region of the tail), but it should lead to well-behaved linear equations.
If the basis is too flexible, we encounter instabilities in 
the ISA functionals, and if the basis is too small, we find that total charge is
not conserved by $0.01$ electrons or more in the minimization of the ISA functionals 
(eqns.~\eqref{eq:stockholder-A-functional} and ~\eqref{eq:stockholder-B-functional}).
Indeed, for a good basis we not only see faster convergence and charge conservation
to $10^{-3}$ electrons or better, but also find that the shape functions 
are positive everywhere with well-defined exponential tails. These criteria can be
used as a means of assessing the quality of the ISA basis sets.

We have found that a reasonable choice for the ISA s-function basis
set is to use an even-tempered set with exponents of the form
$\alpha = 2^n\,\text{a.u.}, n=n_{\text{min}}\dots n_{\text{max}}$, 
where $n_{\text{max}}$ is 5 for hydrogen atoms ($\alpha_{\text{max}}=32.0$)\
and 8 for the heavier atoms ($\alpha_{\text{max}}=256.0$). 
For numerical stability we usually choose $n_{\text{min}}=-3$
($\alpha_{\text{min}}=0.125$). While this choice results in well-behaved
ISA shape functions for most systems, there are cases for which
the 0.125 exponent for hydrogen atoms needs to be omitted, while for silicon
an exponent of $0.0625$ was added. We have termed this set as `ISA/set2'.
With this ISA basis set, the \ISAfunc{A} functional conserves charge to
$10^{-3}$ electrons or better. 

While the above procedure works well for the ISA shape functions \w{a}, 
as mentioned above, it may be that the higher angular momentum functions
in the density-fitting sets also need to be augmented to better model
the ISA atomic densities \R{a}. We are investigating this possibility.

\subsection{Fixing the shape-function tail}
\label{sec:tail-fix}

The ISA shape functions are generally well-behaved in the core density region,
but often exhibit problems in the tail. As mentioned above, negative
terms in the expansion can lead to catastrophic instabilities and need to be
controlled. Furthermore, as there is very little weight given to the 
small densities of the tail region, solutions can easily emerge which have
odd features for low densities (less than about $10^{-5}$ a.u.) Effects of this
order are small, and may have no apparent adverse effect on the overall 
ISA solution, but such behaviour is unsatisfactory.

The first ingredient needed to improve the tail is suggested by the
empirical observation, from calculations on a variety of systems,
that well-converged shape functions tend to exhibit 
an atom-like exponential decay. We therefore require all 
ISA shape functions to decay in this manner;
that is, \w{a}, and hence the ISA atomic density \R{a}, 
should decay exponentially as
\begin{align}
  \wL{a}(\rr) = A_{a} \exp{(-\alpha_{a} |\rr - \RR_{a}|)}, 
    \label{eq:wL}
\end{align}
where $\RR_{a}$ is the centre of atom $a$, and $A_{a}$ and $\alpha_{a}$ are
constants yet to be determined.
We note here that although the more accurate decay of an atomic density
is $A r^{\beta} \exp{(-\alpha r)}$,
we have found that the additional polynomial factor does not appear to be
important, and in the interest of simplicity it is omitted.

We now define the corrected shape function \wT{a} as
\begin{align}
  \wT{a}(\rr) = 
    \begin{cases}
      \w{a}(\rr)  & \text{if} \ |\rr| \leq r^a_0 \\
      \wL{a}(\rr) & \text{otherwise}.
    \end{cases}
    \label{eq:tail-fix}
\end{align}
Here $r^a_0$ is a distance up to which we may expect the uncorrected shape function
$\w{a}(\rr)$ to be reliable. We typically take this to be a fixed multiple 
(usually $1.5$) of the Slater radius \cite{Slater64} of the atom $a$,
though in principle this radius could be determined self-consistently by
examining the manner in which \w{a} decays.
The constants $A_{a}$ and $\alpha_{a}$ in \wL{a} are determined by requiring
continuity at $r_0$ and by ensuring that the charge contained in 
\wT{a} is identical with the charge in \w{a}. Both conditions can be 
enforced analytically. The charge conservation condition is necessary;
without it the corrected shape functions \wT{a} alter the site charges
by a small amount at each step, causing a slow divergence of the
ISA iterations.

We now use the corrected shape functions in the ISA functionals
given in eqns.~\eqref{eq:stockholder-A-functional} and 
\eqref{eq:stockholder-B-functional}.
Notice that \wT{a} is not an expansion in GTOs, but, due to its piece-wise
continuous form, must instead be defined numerically on a grid. 
Consequently, if used in the functional \ISAfunc{A}, this
functional must use the overlap norm, as it is not practical to evaluate
a six dimensional coulomb integral using grids.

To further stabilize the ISA atomic tails we increase the weights given to the
tail by using, instead of the overlap norm, the following tail-weighted norm:
\begin{align}
  \| f \|_{\rm tail} = 
     \int f(\rr) \; \exp(+\epsilon |\rr - \RR_{a}|^2) \; d\rr,
     \label{eq:tail-weighting}
\end{align}
where $\epsilon$ is a positive number that must be less than twice the 
smallest exponent in the basis set so as to ensure integrability.
We apply this weight to the s-function block only.

There is a degree of self-consistency in this process as all the parameters
in the tail correction are updated at each iteration.
Additionally, the shape function tail-weighting is only applied
when the shape function tails have been determined to be sufficiently stabilized.
Finally, both the tail correction and the additional weighting can 
be removed in the final iterations if a fully self-consistent ISA solution is
required.

\subsection{A Robust ISA algorithm}
\label{ssec:DF-ISA-algorithm}

The improvements described above significantly improve the stability and
accuracy of the basis-space ISA (BS-ISA) procedure, but we still see small
charge violations (of the order of $10^{-3}$ electrons)
when minimizing the functionals \ISAfunc{A/B}.
These are very likely linked to the still 
insufficiently flexible ISA basis sets and possibly also to the
nature of the fix applied to the shape-function tails. 
While these charge violations are typically small,
we need a method which will
guarantee a good fit to the density while obtaining the best ISA
solution possible within the basis set constraints. This is possible
by simultaneously minimizing the $\Delta^{\rm DF}$ and $\Delta^{\rm Q}$
functionals together with either \ISAfunc{A} or
\ISAfunc{B}. We will define the \ISA functional
using a single parameter $\zeta \in [0,1]$ to control the relative weights
of the density-fitting and ISA functionals as follows:
\begin{align}
  \DFISAfunc{\zeta} = (1 - \zeta) \; \left( \Delta^{\rm DF} + \Delta^{\rm Q} \right) 
  + \zeta \; \Delta^{\rm stock(A/B)}.
         \label{eq:DFISA-functional}
\end{align}
Notice that we have included the charge conservation functional $\Delta^{\rm Q}$ with
the density-fitting functional. This was done primarily for convenience of implementation;
ideally it might be desirable to include $\Delta^{\rm Q}$ without a 
$\zeta$-dependence.
In the discussion that follows, \DFISAfunc{\zeta} will mean the variant using
\ISAfunc{A} as most of our results have been obtained with this choice.

Ideally we would want our results to be independent of the choice of $\zeta$, 
and, indeed, for well-converged systems we will show that the dependence
on $\zeta$ is small.
This parameter controls the off-diagonal blocks that allow basis functions 
on neighbouring sites to be used to model the density at a given site. 
For $\zeta$ near 1, the diagonal blocks are dominant and we obtain 
a solution close to the pure ISA solution, while for $\zeta$ near 0,
the off-diagonal blocks are large and we achieve a more accurate fit to the
total density, though with a relaxation of the ISA atomic densities.
In practice, for a good \ISA basis set, values of $\zeta$ between
$0.1$ and $0.9$ appear to be 
satisfactory, with very little variation in the final results.

We now describe a robust version of the basis-space ISA method:
\begin{enumerate}
\item Initialize the shape functions \wIT{a}{n=0} as described in 
  the na\"{i}ve ISA algorithm presented in 
  sec.~\ref{ssec:naive-ISA-algorithm}.
\item Attempt to determine the corrected form of the shape-function tails
  given in eq.~\eqref{eq:tail-fix}. This is not always possible, and if 
  the parameters of the function \wL{a} are deemed to be
  unphysical, the fix is not attempted.
\item Minimize the \DFISAfunc{\zeta} functional to obtain the ISA atomic 
  densities \R{a}.
\item Update the shape functions \wIT{a}{n+1} using the ISA atomic
  densities \R{a} (eq.~\eqref{eq:w-expansion}).
\item Check convergence by evaluating:
  \begin{align}
    d^{a} = \frac{<\wIT{a}{n}|\wIT{a}{n+1}>}
             {\sqrt{<\wIT{a}{n}|\wIT{a}{n}> <\wIT{a}{n+1}|\wIT{a}{n+1}>}}.
       \label{eq:conv}
  \end{align}
  With a convergence parameter $\epsilon$ (typically $10^{-9}$), 
  we have achieved convergence if $|1 - d^{a}| \leq \epsilon\; \forall\; a$. 
\item If converged, exit.
\item If the shape functions are deemed to be
  sufficiently stabilized, turn on the additional tail weighting
  (eq.~\eqref{eq:tail-weighting}). This is usually done if 
  convergence is attained to $\tilde{\epsilon} = 10^{-5}$.
\item Iterate from step 2.
\end{enumerate}
Because this algorithm combines density-fitting and the \ISA methods, we will 
refer to this as the \DFISA algorithm.

We mention here that there are many variants of the \DFISA algorithm:
We could, for example, use the \DFISAfunc{\zeta} functional with a fixed
value of $\zeta$ until convergence is attained. 
Alternatively, we could minimize \DFISAfunc{\zeta=1.0} so as to converge the
shape functionals, and only then reduce $\zeta$ to fit the total density better. 
This variant has the advantage that $\zeta = 1.0$ corresponds to minimizing 
the \ISAfunc{A} functional only, and this can be made to scale linearly with
the number of atoms. Consequently, this approach may be better suited to
large systems. Once convergence has been attained, 
with $\zeta \lt 1.0$, the off-diagonal blocks in the DF-ISA matrices
are non-zero and allow the solution to relax to fit the total
density better, though with a slight degradation of the ISA solution.

\subsection{Relation to the GISA variant}
\label{sec:GISA}

As noted above, Verstraelen \etal \cite{VerstraelenAvanSW12} have recently described an
analogous approach that they call the Gaussian-ISA, or GISA, method.
It is formally exactly the same as the Lillestolen--Wheatley ISA except that the
shape functions are described using an expansion in a series of 
s-functions as proposed here.
Our proposal differs from the GISA method in important ways: firstly,
we use a far more complete set of s-functions which are essential 
for adequate convergence of the shape-function tails.
We find that this flexibility is needed
to ensure that functional \ISAfunc{A} conserves charge to $10^{-3}$ electrons
or better (without application of the global charge conservation constraint).
As we have noted above, the better the basis set, the better charge 
is conserved, as the functional is able to satisfy the 
ISA conditions and reproduce the total density simultaneously. 
With the GISA basis sets we see charge violations of between 0.01 (formamide)
and 0.2 (benzene) while the ISA/set2 results in significantly smaller charge
violations of just 0.001 to 0.003 electrons, clearly indicating that the 
ISA/set2 basis possesses the variational flexibility needed to accurately
describe the ISA shape functions. 

\begin{figure}[h]
  \includegraphics[viewport=0 40 550 340,clip,width=0.45\textwidth]{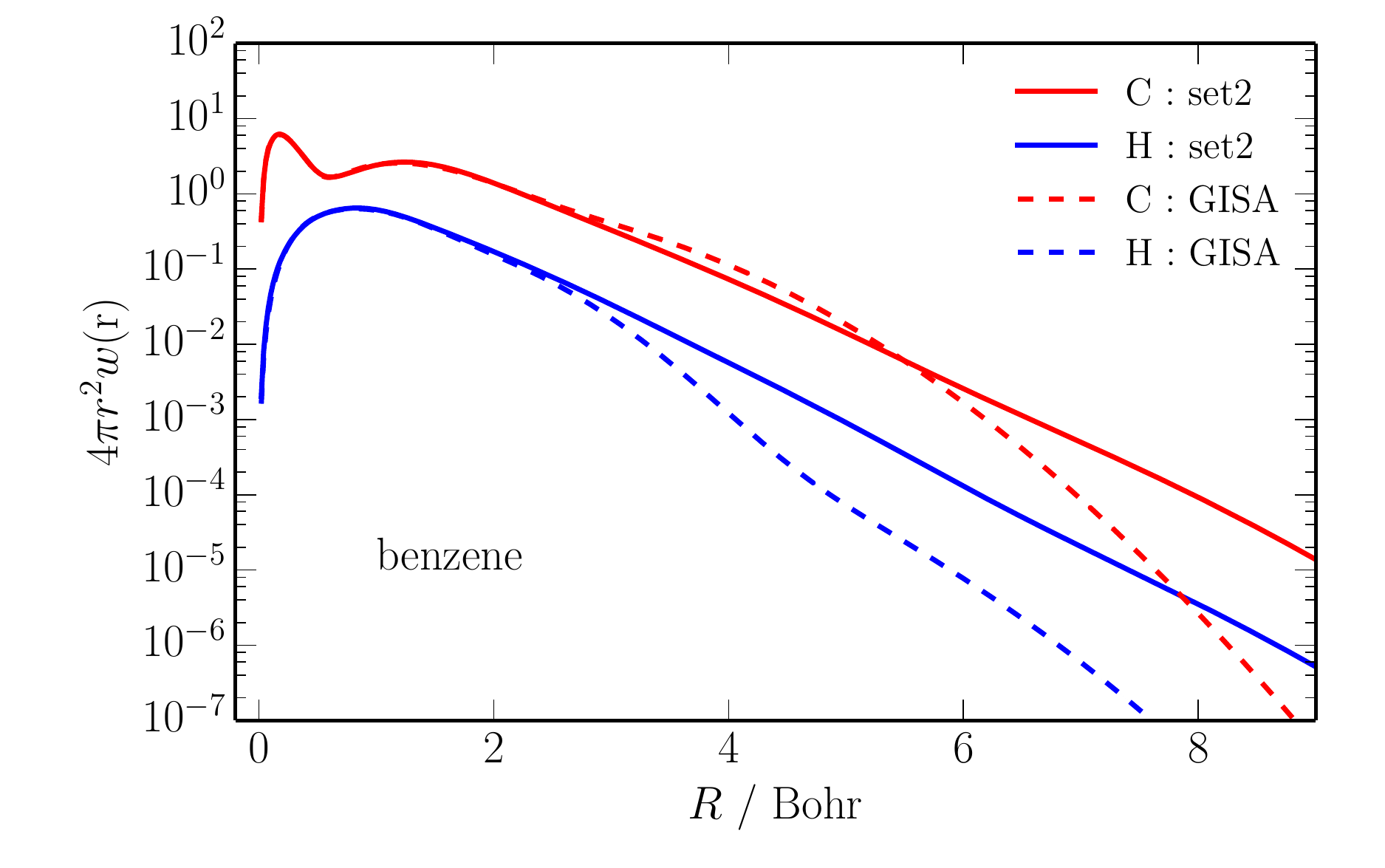}
  \caption[GISA versus set2]{
  Shape functions for the carbon and hydrogen atoms in benzene (aug-cc-pVTZ basis)
  calculated using the GISA basis sets and the ISA/set2 basis set described
  in this paper. In both cases functional \ISAfunc{A} was minimized.
  \label{fig:GISA_vs_set2_Bz}
  }
\end{figure}

Of course, charge conservation is not an issue if the stockholder partitioning
(eq.~\eqref{eq:stockholder}) is performed in real space, as is done
by Verstraelen \etal  and in the original formulation by  Lillestolen and Wheatley.
In that case, as long as the shape functions are positive everywhere and finite 
in extent, no matter how pathological they might otherwise be, charge will
always be conserved. From our experience, and the results of Verstraelen \etal, 
the shape functions obtained using the GISA basis sets
can be reasonable, but we find that they often exhibit pathologies in the way the
tails decay. One such example is illustrated in fig.~\ref{fig:GISA_vs_set2_Bz}
for the benzene molecule (aug-cc-pVTZ basis). The shape functions 
have been obtained by minimizing \ISAfunc{A} using the GISA and ISA/set2 
basis sets with all other parameters the same. The ISA/set2 atomic shapes 
exhibit a clear exponential decay (in the range shown in the figure)
which contrasts with the somewhat erratic and non-exponential decay of
the shape functions obtained using the GISA basis sets.

Finally we point out that our approach results in linear equations that are readily 
suited for applications to large molecular systems and that, because our 
ISA s-function basis sets are created using a simple algorithm, the basis sets
can be extended to other atomic systems with relatively little effort.

\subsection{\DFISA numerical details and implementation}
\label{ssec:DFISA-details}

Minimizing the density-fitting and charge-conservation functionals,
eqns.~\eqref{eq:DF-functional} and \eqref{eq:Q-functional}, leads to
the following linear equations:
\begin{align}
   S^{\rm DF}_{k,k'} d_{k'} = T^{\rm DF}_{k},
    \label{eq:DF-Q-lineq}
\end{align}
where $\mathbf{d}$ is the coefficient vector for the density expansion
given in eq.~\eqref{eq:rho-expansion}, and the L.H.S.\ matrix $\mathbf{S}^{\rm DF}$ 
and R.H.S.\ vector $\mathbf{T}^{\rm DF}$ are defined as
\begin{align}
  S^{\rm DF}_{k,k'} & = < k||k'> + \lambda I_k I_{k'} \label{eq:DF-Q-Smat} \\
  T^{\rm DF}_{k}    & = <\rho || k> + \lambda N I_{k}, \label{eq:DF-Q-Tmat} 
\end{align}
where $<k||k'>$ signifies the coulomb integral of the basis functions 
$\chi_{k}$ and $\chi_{k'}$, and $I_k = \int \chi_k(\rr) \; d\rr$.
The $\mathbf{S}$ matrix is order $N \times N$, where $N$ is the number
of auxiliary basis functions in the system. Therefore the computational
cost of solving eq.~\eqref{eq:DF-Q-lineq} scales as $\order{N^3}$.
On the other hand, the functionals \ISAfunc{A} and \ISAfunc{B}
are both expressed as the sum over sites of functionals that depend on the 
auxiliary basis functions located on the site only. This allows us to 
perform the minimization of these functionals piece-wise, with \order{N} 
computational cost, by minimizing one block at a time. For example,
\ISAfunc{A} can be written as
\begin{align}
  \ISAfunc{A} &= \sum_{a} \left\| 
         \biggl( \R{a} - \rho \frac{\w{a}}{\sum_b \w{b}} \biggr)^2
                                 \right\|, \nonumber \\
              &= \sum_{a} \ISAfunc{A}^{a}.
  \label{eq:stockholder-A-functional-blocks}
\end{align}
The minimization of $\ISAfunc{A}^{a}$ leads to the equations
\begin{align}
   \tilde{S}^{a}_{k,k'} d^{a}_{k'} = \tilde{T}^{a}_{k},
    \label{eq:ISA-A-lineq}
\end{align}
where the superscript $a$ indicates that the quantities defined here 
depend only on functions centered at site $a$, and 
\begin{align}
  \tilde{S}^{a}_{k,k'} &= < k||k' > \label{eq:ISA-A-Smat} \\
  \tilde{T}^{a}_{k}    &= \int \chi^{a}_{k}(\rr) \rho(\rr) 
                               \frac{\w{a}(\rr)}{\sum_b \w{b}(\rr)} \; d\rr.
             \label{eq:ISA-A-Tmat}
\end{align}
Notice that when considered as a matrix over all sites $a$, the $\tilde{\mathbf{S}}$ matrix
is block-diagonal. 

For each site $a$, the solution of eq.~\eqref{eq:ISA-A-lineq} involves a computational
cost of order $\order{N^0}$. As there are \order{N} sites, the total computational
cost of minimizing \ISAfunc{A} scales as \order{N}.
However to achieve this linear scaling the $\tilde{T}^{a}_{k}$ integrals must be 
calculated using locality. In \CamCASP we did this as follows:
\begin{itemize}
  \item Neighbours are defined for every site. A site $b$ is considered a 
  neighbour of site $a$ if the overlap integral of the most diffuse auxiliary basis
  function (all functions are treated as s-functions for this purpose)
  of sites $a$ and $b$ exceeds a specified threshold. This is a reasonable definition
  as density-fitting is used for all quantities in the \CamCASP program.
  \item The integration grid used in eq.~\eqref{eq:ISA-A-Tmat} is constructed from
  the atom grids of site $a$ and neighbouring sites only.
  \item Likewise, the density-fitted molecular density $\rho$ and pro-molecular
  density $\sum_b \w{b}(\rr)$ are evaluated using auxiliary basis functions located on
  the site $a$ and its neighbours only.
\end{itemize}
With these considerations, $\tilde{T}^{a}_{k}$ can be evaluated with computational 
cost of order $\order{N^0}$.

The linear equations for the \DFISA functional in eq.~\eqref{eq:DFISA-functional}
can be obtained by constructing the ISA equations for the full molecular system
from the atomic site equations in eq.~\eqref{eq:ISA-A-lineq}, and 
combining this set of equations with the density-fitting equations
in eq.~\eqref{eq:DF-Q-lineq} to give
\begin{align}
   \mathcal{S}_{k,k'} d_{k'} = \mathcal{T}_{k},
     \label{eq:DF-ISA-A-lineq}
\end{align}
where if $k \in a$ and $k' \in a'$ then
\begin{align}
  \mathcal{S}_{k,k'} &= (1-\zeta) S^{\rm DF}_{k,k'} + \zeta \left(
                      \sum_{b} \delta_{ba} \delta_{ba'} \tilde{S}^{b}_{k,k'}
                                     \right) \\
  \mathcal{T}_{k}    &= (1-\zeta) T^{\rm DF}_{k} + \zeta \left(
                      \sum_{b} \delta_{ba}  \tilde{T}^{a}_{k}
                                     \right).
\end{align}
These equations can be solved exactly like the standard density-fitting equations.

\section{\DFISA shape functions and convergence}
\label{sec:w_and_conv}

The \DFISA calculations reported in this paper 
have been performed using atomic densities calculated using the 
PBE0 \cite{AdamoB99a} functional, asymptotically corrected using the
Fermi--Amaldi \cite{FermiA34} correction and the Tozer \& Handy \cite{TozerH98}
splicing scheme. Density functional calculations have been performed
using the \DALTON program \cite{DALTON2} using a patch included with the
\SAPT \cite{SAPT2008} program. Unless otherwise specified, we have used the
Dunning aug-cc-pVQZ basis \cite{Dunning89,KendallDH92} for density 
functional calculations. Vertical ionisation potentials needed for the 
asymptotic correction have either been calculated using the 
$\Delta$-DFT algorithm or have been taken from the NIST Chemistry Web-book 
\cite{NIST_lias-long}.

The \DFISA algorithm described here has been implemented in a pre-release
version of the \CamCASP program \cite{CamCASP} and is available from the authors upon
request. Iso-density maps reported in this paper have also been calculated 
with \CamCASP and are displayed using the \ORIENT program \cite{Orient}.
All multipole models have been calculated using the \CamCASP program.

We have investigated a number of systems including H$_2$, H$_2$O, CH$_4$,
CCl$_4$, NH$_3$, LiF, H$_2$CO, CO, CO$_2$, pyridine (d-aug-cc-pVTZ basis),
benzene (aug-cc-pVTZ basis), formamide, HF and C$_{10}$N$_2$H$_{13}$
(aug-cc-pVDZ basis), but will report only a subset of the data in this paper.

\subsection{Convergence}
\label{sec:conv}
The biggest problem associated with the ISA algorithm has been its poor
convergence properties. Real-space algorithms can take more than a 
thousand iterations to converge or may not converge at all
\cite{VerstraelenAvanSW12}. The GISA algorithm of Verstraelen \etal
fares far better with algorithmic convergence in 140 iterations or so.
However, due to the restricted variational flexibility of the GISA 
basis sets, this is not true convergence, as we have pointed out above.

\begin{figure*}[]
  \includegraphics[viewport=0 40 550 340,clip,width=0.45\textwidth]{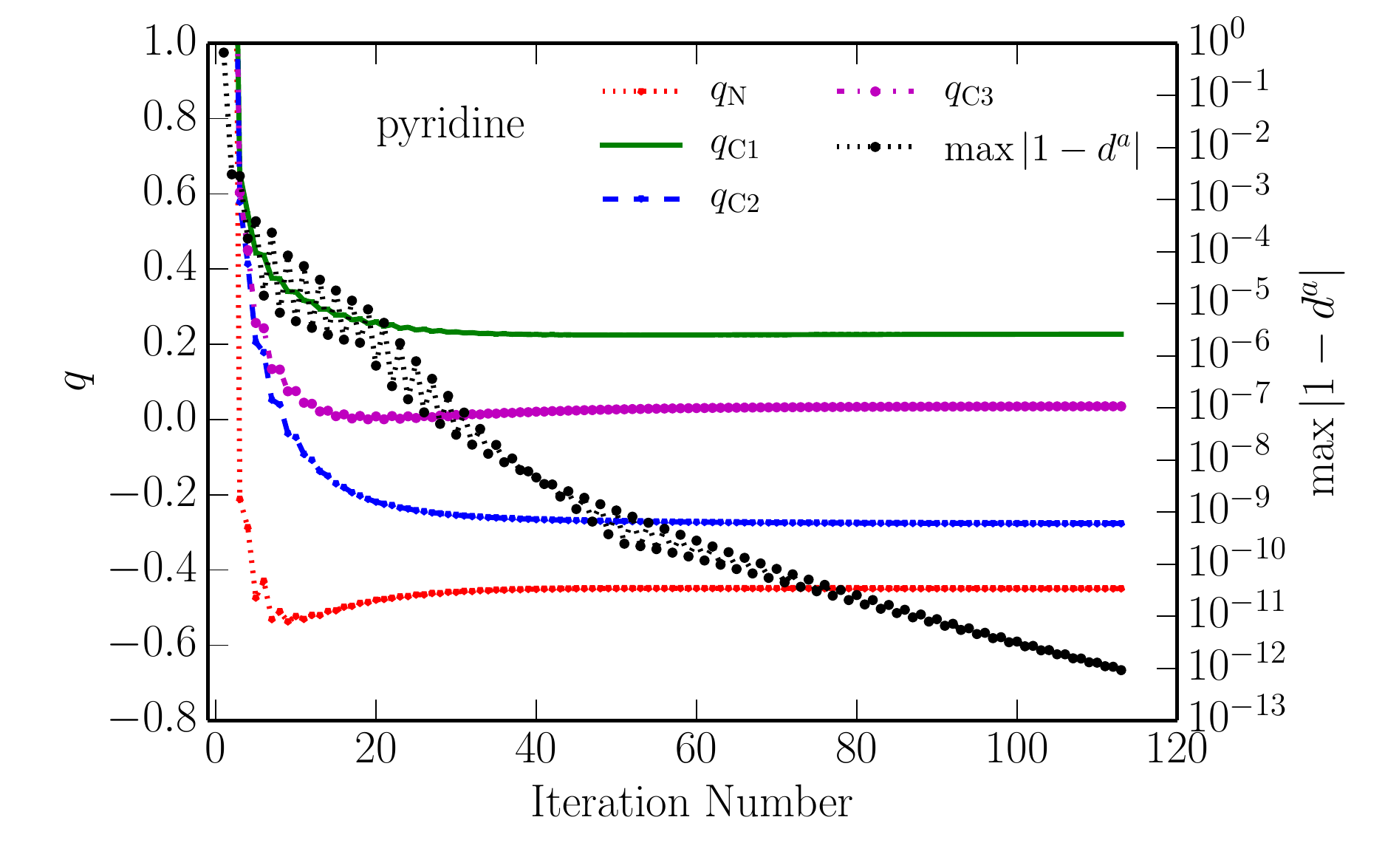}
  \includegraphics[viewport=0 40 550 340,clip,width=0.45\textwidth]{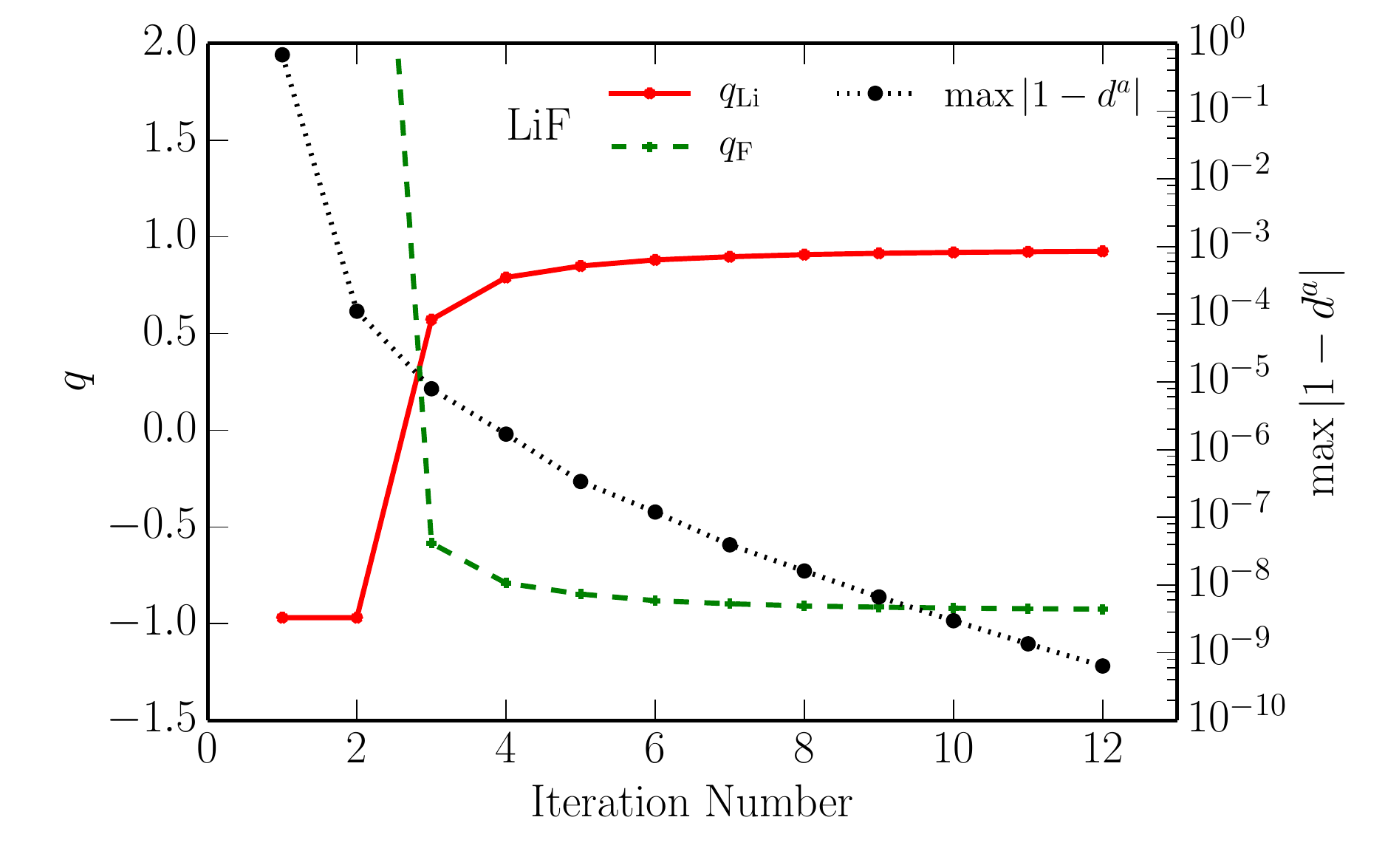}
  \includegraphics[viewport=0  0 550 340,clip,width=0.45\textwidth]{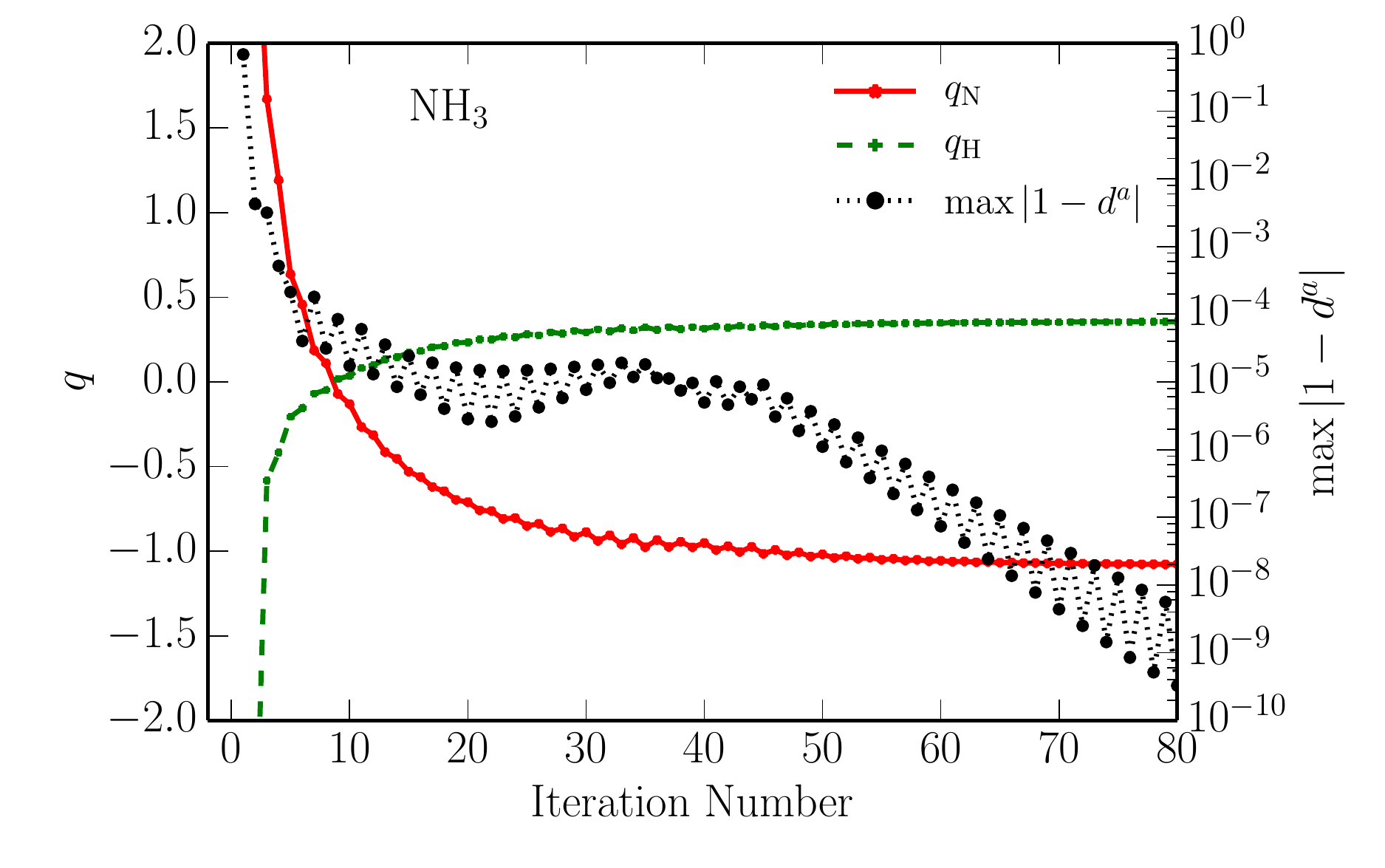}
  \includegraphics[viewport=0  0 550 340,clip,width=0.45\textwidth]{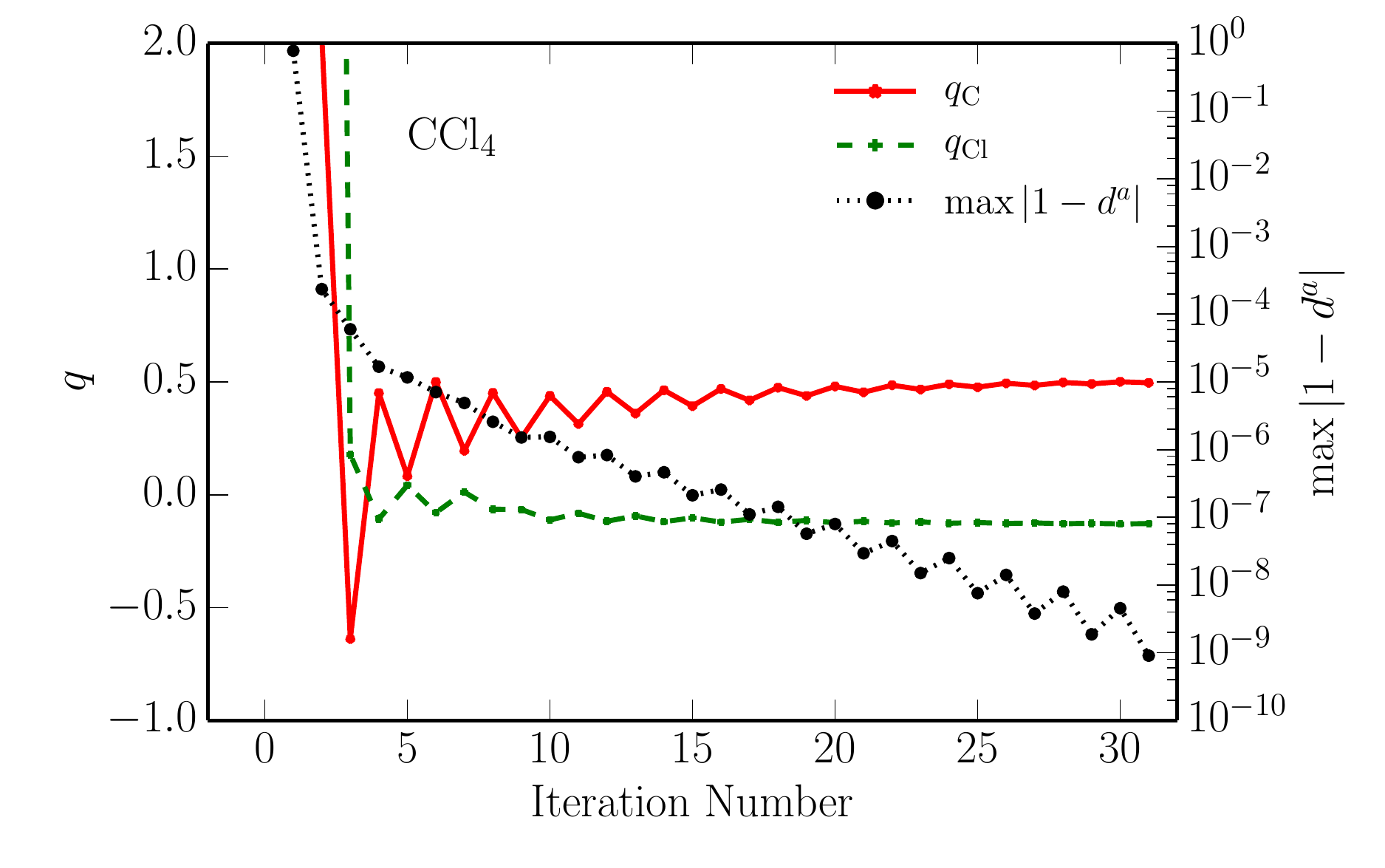}
  \caption[Convergence patterns]{
  \ISA convergence patterns for pyridine, LiF, NH$_3$ and CCl$_4$.
  We have plotted the AIM charges (left $y$-axis) and convergence parameter 
  $\max|1-d^a|$ (right $y$-axis; see eq.~\eqref{eq:conv}) against iteration number.
  Most calculations use a convergence threshold of $10^{-9}$, but for pyridine we
  have reduced the threshold to $10^{-12}$ to better illustrate the convergence
  properties of the \ISA algorithm.
  The pyridine density was calculated using the d-aug-cc-pVTZ basis while for the
  the other molecules the aug-cc-pVQZ basis was used.
  \label{fig:conv}
  }
\end{figure*}

There are no such issues with the \DFISA algorithm, which we have found
to converge in at
most 80 iterations and sometimes as few as 10, without any convergence
acceleration techniques that might be applicable to the algorithm.
There is no apparent effect of system size on the number of iterations
required for convergence, though we have noticed that basis set 
improvements can lead to even faster convergence, and conversely,
a small or unbalanced basis can lead to poor convergence.
In fig.~\ref{fig:conv} we display convergence patterns for the \DFISA
algorithm for a representative sample of the systems we have 
investigated. 
Our normal convergence criterion is that all shape functions converge
to $10^{-9}$ or better using eq.~\eqref{eq:conv}, but for illustrative
purposes we have chosen a threshold of $10^{-12}$ for the pyridine 
molecule (density obtained with the d-aug-cc-pVTZ basis).
Plotted together with $\max|1-d^a|$ are the ISA charges of the heavier 
atoms of pyridine. The hydrogen atoms are omitted for clarity.
The charges can be seen to converge very smoothly, but $\max|1-d^a|$
exhibits oscillations that die off just before the 40$^{th}$ iteration,
only to re-appear and subsequently die off again. This is quite a
common occurrence; it can be seen for the NH$_3$ system too,
% (though in that case
% we have not included data beyond the usual convergence criterion) 
and we have no explanation for this behaviour.

The LiF molecule shows the fastest convergence of any we have studied, with smooth
and rapid convergence in 12 iterations. While the same is true for CCl$_4$, here
we observed oscillations in the ISA site charges. These die off by iteration 25.
The convergence patterns for the other systems we have studied fall into one of these
four categories and are not presented here. 

Notice that in all cases, past a threshold, convergence is exponentially fast with
iteration number. This seems to be true in general. Additionally, there is no
apparent relation between rate of convergence and system size: the relatively
small ammonia molecule took 80 iterations to converge, but the largest system we 
have investigated (25 atoms) took 53 iterations. This is particularly promising as this
is a desirable property for large applications.

\subsection{Shape Functions}
\label{sec:shape-functions}

\begin{figure*}
  \includegraphics[viewport=0 40 550 340, clip, width=0.45\textwidth]{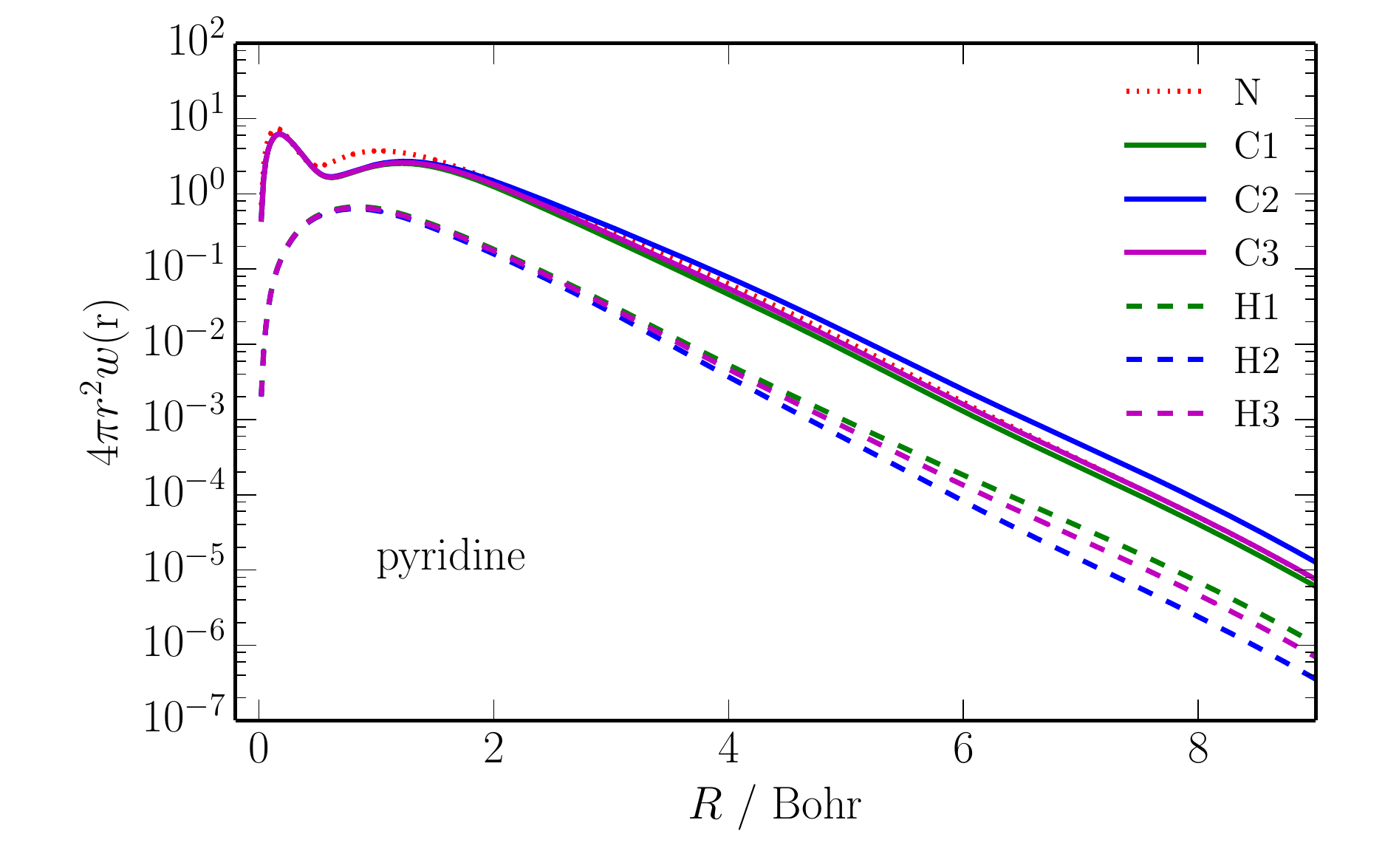}
  \includegraphics[viewport=0 40 550 340, clip, width=0.45\textwidth]{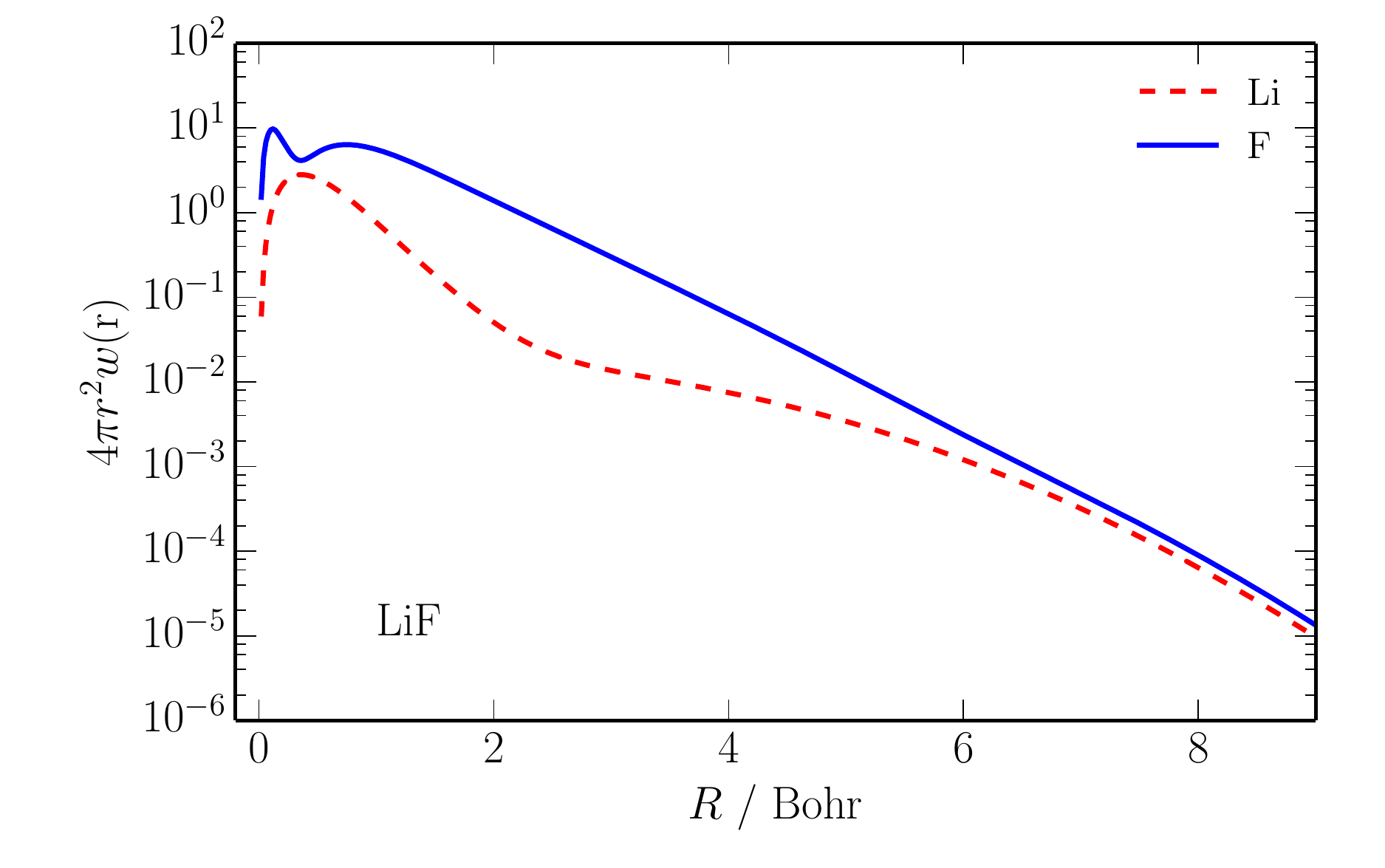}
  \includegraphics[viewport=0  0 550 340, clip, width=0.45\textwidth]{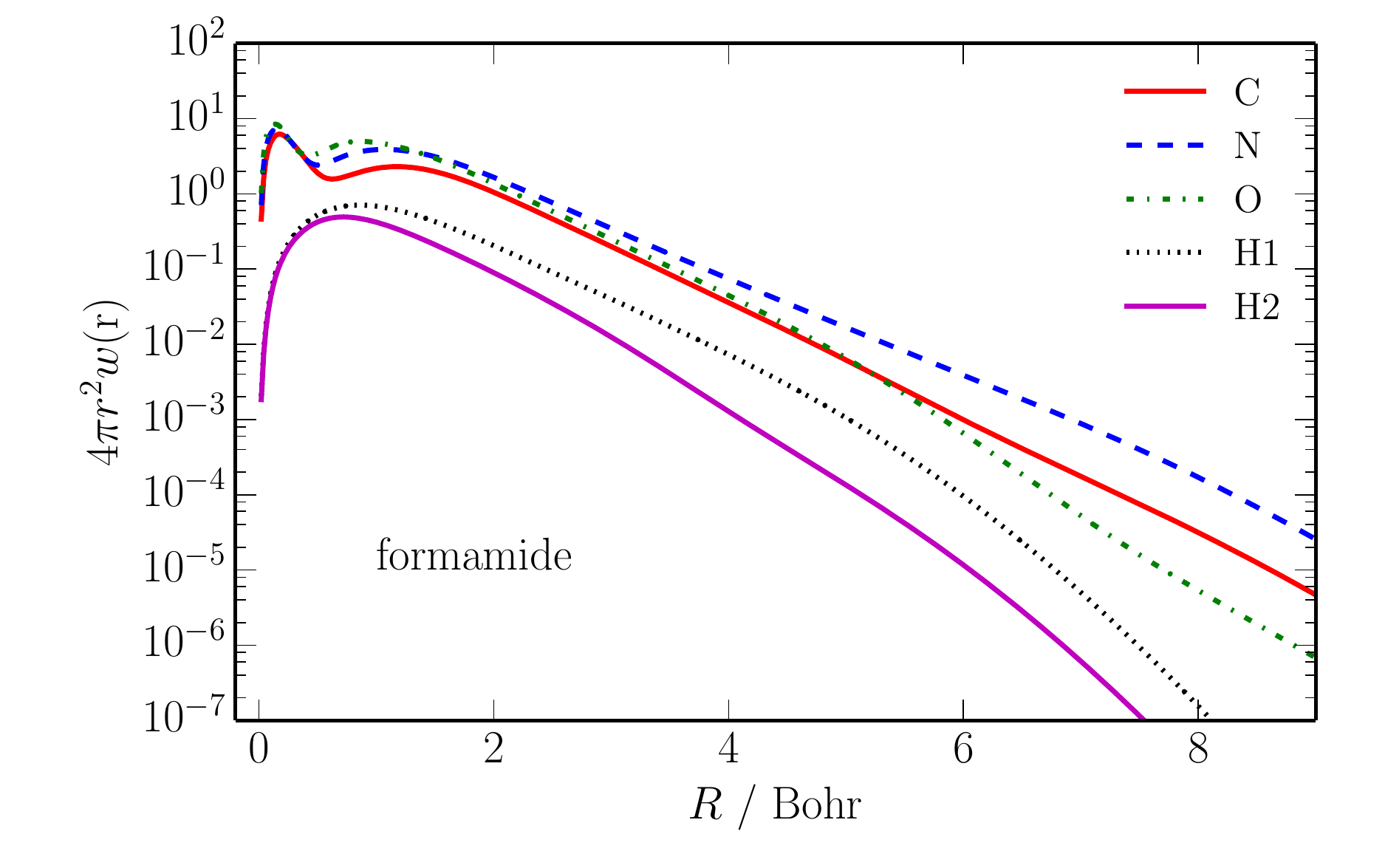}
  \includegraphics[viewport=0  0 550 340, clip, width=0.45\textwidth]{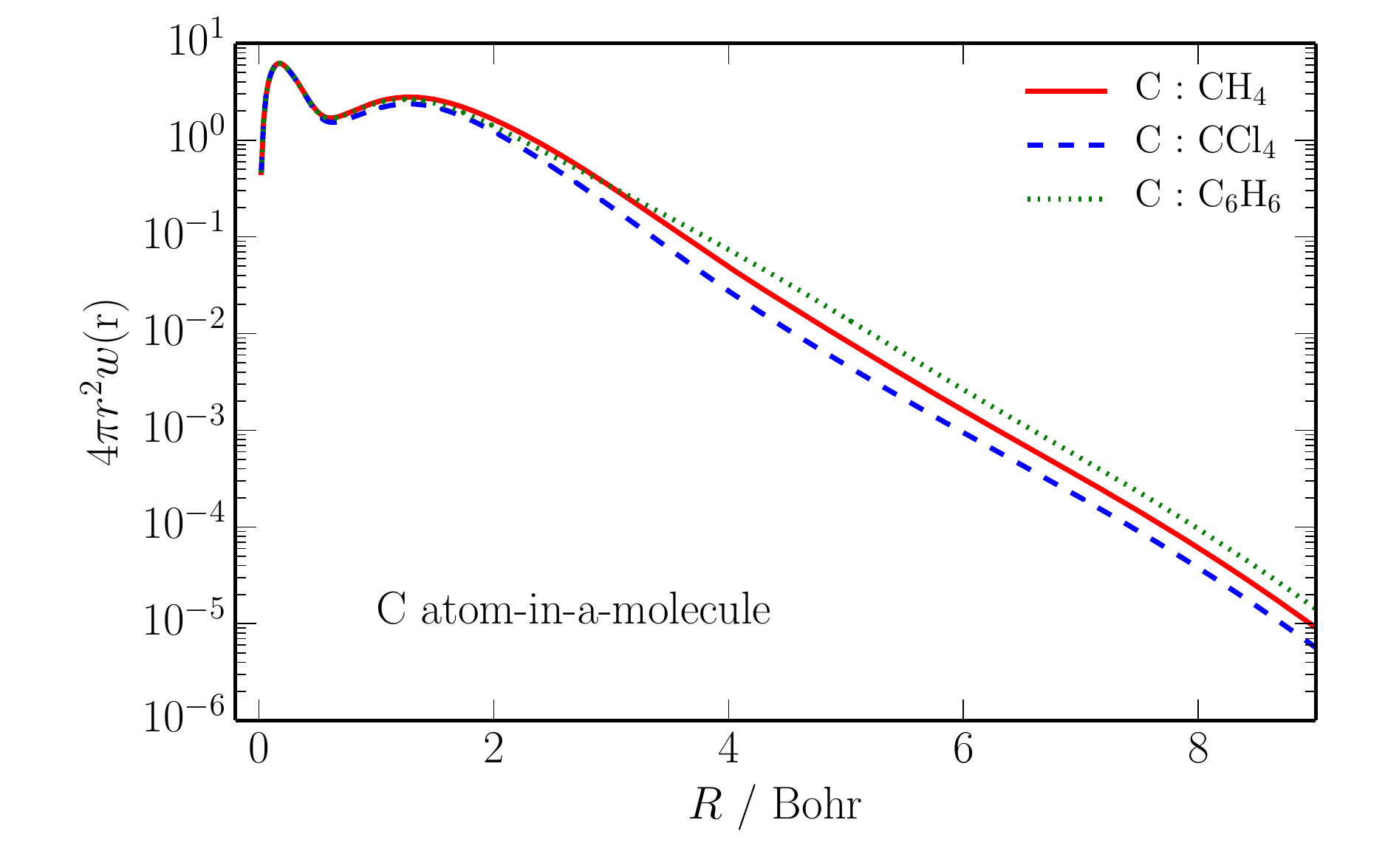}
  \caption[Shape functions]{
  \DFISA shape functions for atoms in pyridine, LiF and formamide.
  We have plotted $4\pi r^2 w(r)$ to better illustrate the shell structure. 
  The aug-cc-pVQZ basis was used to calculate the density, except for
  pyridine, where d-aug-cc-pVTZ was used.
  The \DFISA calculation was performed using the aQZ/set2 basis, but for formamide
  we had to limit the $s$-functions on the hydrogen atoms to a smallest exponent 
  of $0.25$ a.u.\ The amide hydrogen atoms in formamide are very
  similar, and for clarity only one is shown.
  In the last panel carbon shape functions are compared for carbon atoms in
  benzene, methane and carbon tetrachloride.
  All shape functions were obtained using the \DFISA algorithm with $\zeta=0.9$. 
  \label{fig:w-various}
  }
\end{figure*}

In fig.~\ref{fig:w-various} we report shape functions for the atoms in the 
pyridine, formamide, LiF and CCl$_4$ systems. Rather than plot $w(r)$ directly,
we have plotted $4\pi r^2 w(r)$ to better illustrate the shell structure of the
atoms in these systems. All shape functions have been calculated with the
\DFISA algorithm with $\zeta=0.9$. This value of $\zeta$ was chosen as shape functions
are generally better behaved for $\zeta$ closer to $1$, when the \ISAfunc{B} is
dominant. 

The pyridine molecule illustrates the success of the \DFISA method. All atomic
shape functions are well-behaved, with clear exponential tails. This is not always
the case. For the formamide molecule we were able to obtain shape functions that
were positive everywhere only after eliminating the most diffuse ($\alpha=0.125$) 
s-function from the ISA basis set for the hydrogen atoms. This explains why 
the H1 and H2 hydrogen shape functions of formamide decay quickly past 6 Bohr.
Despite these changes, the shape function for the oxygen AIM appears to be 
somewhat spurious past 6 Bohr. Very small changes to the shape function expansions
are responsible for this kind of non-exponential decay, and they occur
in regions where the density is so small that, even with the schemes
described in sec.~\ref{sec:tail-fix}, it can sometimes be difficult to
control the behaviour of the shape functions.

The shape functions for the LiF molecule clearly pick out the substantial
differences between the Li and F atoms in LiF. In principle, the lithium AIM should determine the
density decay of the system and, indeed, it seems to have the slower decay until 
about 6 Bohr when the decay of both atoms become similar. This is a 
consequence of the basis set used. Both atoms use the same ISA s-function
basis set, so at sufficiently long-range their shape functions must 
decay in a similar manner.

Also shown in fig.~\ref{fig:w-various} are shape functions for carbon 
atoms in different molecules. The ISA charges on these carbon atoms are
$-0.11$ (benzene), $-0.42$ (methane), and $+0.51$ (CCl$_4$). This is
what would be expected from the electronegativity differences of the bonding 
atoms. The \DFISA algorithm correctly shows that the carbon AIM density
in methane is more diffuse than the carbon in carbon tetrachloride. 
The carbon atom in benzene is more diffuse than the methane and CCl$_4$
carbons due to the planar nature of the benzene molecule.

\section{Multipoles}
\label{sec:multipoles}

\begin{table}
\newcolumntype{d}{D{.}{.}{2}}
%  Table of differences between various multipole models and a SAPT
%  scan of the electrostatic potential over a surface at twice the
%  van der Waals radii. Values are in millivolt.
\begin{tabular}{ldddddd}
\toprule
\multicolumn{1}{r}{Rank:} & 0 & 1 & 2 & 3 & 4 & 5\\
\midrule
\multicolumn{5}{l}{\emph{DMA0}}\\
$<\Delta>$           &  -1.63 &   0.44 &   0.95 &   0.46 &  0.39 &  0.36 \\
$\sigma_{\Delta}$    & 205.45 &  40.36 &  14.93 &   3.73 &  1.03 &  0.79 \\
% max     & 325.97 & 118.16 &  47.27 &  10.43 &  2.47 &  1.97 \\
% r.m.s.  & 205.44 &  40.36 &  14.96 &   3.76 &  1.10 &  0.87 \\
% min     &-355.09 & -93.96 & -32.32 & -11.56 & -2.62 & -2.13 \\
%
\multicolumn{5}{l}{\emph{DMA4}}\\
$<\Delta>$           &   -1.15 &   -4.29 &  -0.55 &  0.02 &  0.07 &  0.08 \\
$\sigma_{\Delta}$    &   53.32 &   98.98 &  11.37 &  2.48 &  1.20 &  0.82 \\
% max     &  104.09 &  231.53 &  24.81 &  5.26 &  3.75 &  1.98 \\
% r.m.s.  &   53.33 &   99.06 &  11.38 &  2.48 &  1.20 &  0.82 \\
% min     & -147.65 & -200.00 & -29.78 & -7.48 & -3.63 & -2.22 \\
%
% \multicolumn{5}{l}{\emph{DMA0, refined using} \textsc{Mulfit}}\\
% $<\Delta>$           &    6.33 &    9.16 &   9.46 \\
% $\sigma_{\Delta}$    &  131.28 &   49.93 &  11.43 \\
% max     &  204.56 &   98.16 &  22.17 \\
% r.m.s.  &  131.43 &   50.75 &  14.84 \\
% min     & -455.25 & -149.35 & -44.38 \\
% \multicolumn{5}{l}{\emph{ISA}}\\
% \multicolumn{5}{l}{(zeta = 0.9)}\\
%  $<\Delta>$          &    6.35 &    8.13 &   8.89 &   8.51 &   8.42 \\
% $\sigma_{\Delta}$    &   66.82 &   33.86 &  33.68 &  17.12 &  17.22 \\
% max     &  102.08 &   90.71 &  62.57 &  23.21 &  20.21 \\
% r.m.s.  &   67.12 &   34.82 &  34.83 &  19.12 &  19.17 \\
% min     & -150.00 & -106.73 & -96.26 & -58.04 & -67.80 \\
% \multicolumn{5}{l}{(zeta = 0.1)}\\
%  $<\Delta>$          &   -0.73 &   -0.23 &  -0.22 &  0.30 &   0.30 \\
% $\sigma_{\Delta}$    &   35.90 &   17.60 &  14.67 &  3.35 &   2.92 \\
% \multicolumn{5}{l}{(Latest DF+ISA)}\\
%  $<\Delta>$          & -0.87 & -0.57 & -0.15 &  0.33 &  0.32 \\
% $\sigma_{\Delta}$    & 32.32 & 19.67 & 10.81 &  2.65 &  2.18 \\
\multicolumn{5}{l}{\emph{\DFISA}, $\zeta = 0.1$}\\
 $<\Delta>$          & -0.89 & -0.49 & -0.41 &  0.21 &  0.20 \\
$\sigma_{\Delta}$    & 33.91 & 18.62 & 14.52 &  3.62 &  3.14 \\
\bottomrule
\end{tabular}
\caption{Table of differences between the electrostatic potential
calculated using \CamCASP (SAPT-DFT) and various multipole
models over a surface at twice the van der Waals radii of the
atoms in the formamide molecule. Values are in millivolt (mV).
$<\Delta> = < \Eelst - \EelstDM>$ is the mean difference and
$\sigma_{\Delta}$  is the standard deviation.
See the text for details.
}
\label{table-diffs2.0}
\end{table}

\begin{table}
\newcolumntype{d}{D{.}{.}{1}}
%  Table of differences between various multipole models and a SAPT
%  scan of the electrostatic potential over a surface at 1.5 times the
%  van der Waals radii. Values are in millivolt.
\begin{tabular}{ldddddd}
\toprule
\multicolumn{1}{r}{Rank:} & 0 & 1 & 2 & 3 & 4 & 5\\
\midrule
\multicolumn{5}{l}{\emph{DMA0}}\\
 $<\Delta>$          &  4.0   &  10.5 &  12.3 &  9.8  &  9.6 &  9.5  \\
$\sigma_{\Delta}$    &  322.9 &  88.5 &  42.5 &  11.5 &  4.9 &  4.6  \\
% max     &  492. &  216. &  109. &  17. &   4. &   1.2 \\
% r.m.s.  &  323. &  89.  &   44. &  15. &  11. &   9.5 \\
% min     & -579. & -233. &  -99. & -46. & -22. & -24.1 \\
%
\multicolumn{5}{l}{\emph{DMA4}}\\
 $<\Delta>$          &    5.8 &    4.9 &    6.2 &   8.7 &    8.9 &    9.0  \\
$\sigma_{\Delta}$    &  108.2 &  199.3 &   32.2 &  10.5 &    6.9 &    5.3  \\
% max     &  255. &  472. &  72. &  22. &  15. &   5.2 \\
% r.m.s.  &  108. &  199. &  33. &  14. &  11. &  10.5 \\
% min     & -294. & -401. & -71. & -38. & -25. & -26.5 \\
%
% \multicolumn{5}{l}{\emph{DMA0, refined using} \textsc{Mulfit}}\\
%  $<\Delta>$          &    6.3 &   9.2 &   9.5 \\
% $\sigma_{\Delta}$    &  131.3 &  49.9 &  11.4 \\
% max     &  204. &   98. &  22. \\
% r.m.s.  &  131. &   51. &  15. \\
% min     & -455. & -149. & -44. \\
% \multicolumn{5}{l}{\emph{ISA}}\\
% \multicolumn{5}{l}{(zeta = 0.9)}\\
%  $<\Delta>$          &   6.4 &   8.1 &   8.9 &   8.5 &   8.4 \\
% $\sigma_{\Delta}$    &  67.1 &  33.9 &  33.7 &  17.1 &  17.2 \\
% max     &  102. &   91. &  63. &  23. &  20. \\
% r.m.s.  &   67. &   35. &  35. &  19. &  19. \\
% min     & -150. & -106. & -96. & -58. & -68. \\
% \multicolumn{5}{l}{(zeta = 0.1)}\\
%  $<\Delta>$          &   6.6 &   8.0 &   8.1 &   9.2 &  9.2 \\
% $\sigma_{\Delta}$    &  65.0 &  37.4 &  34.5 &  10.0 &  8.6 \\
% \multicolumn{5}{l}{(Latest DF+ISA)}\\
%  $<\Delta>$          &   6.0 &   6.9 &   8.1 &  9.3 &  9.3 \\
% $\sigma_{\Delta}$    &  62.9 &  40.7 &  25.6 &  8.3 &  6.7 \\
\multicolumn{5}{l}{\emph{\DFISA}, $\zeta = 0.1$}\\
 $<\Delta>$          &   6.0 &   7.2 &   7.3 &   8.9 &  8.9 \\
$\sigma_{\Delta}$    &  64.2 &  38.9 &  34.2 &  10.0 &  8.3 \\
\bottomrule
\end{tabular}
\caption{
Table of differences between the electrostatic potential
calculated using \CamCASP (SAPT-DFT) and various multipole
models over a surface at $1.5$ times the van der Waals radii of the
atoms in the formamide molecule. Values are in millivolt (mV).
See caption to table \ref{table-diffs2.0} for details.
}
\label{table-diffs1.5}
\end{table}

\begin{table}
\newcolumntype{d}{D{.}{.}{1}}
%  Table of differences between various multipole models and a SAPT
%  scan of the electrostatic potential over an isodensity surface at an
%  electron density of 0.001 a.u.. Values are in millivolt.
\begin{tabular}{ldddd}
\toprule
\multicolumn{1}{r}{Rank:} & 0 & 1 & 2 & 4\\
\midrule
\multicolumn{5}{l}{\emph{DMA0}}\\
 $<\Delta>$          & +78.8 & +92.2 & +95.4 & +86.0 \\
$\sigma_{\Delta}$    & 496.5 & 200.0 & 121.3 &  30.0 \\
\multicolumn{5}{l}{\emph{DMA4}}\\
 $<\Delta>$          & +82.9 & +44.6 & +74.6 & +84.0 \\
$\sigma_{\Delta}$    & 243.6 & 408.4 & 114.0 &  47.7 \\
\multicolumn{5}{l}{\emph{\DFISA}, $\zeta = 0.1$}\\
 $<\Delta>$          & +79.8 & +82.9 & +82.9 & +85.0 \\
$\sigma_{\Delta}$    & 132.7 &  84.0 &  87.1 &  40.4 \\
\bottomrule
\end{tabular}
\caption{
Table of differences between the electrostatic potential
calculated using \CamCASP (SAPT-DFT) and various multipole
models for the formamide molecule
over an isodensity surface at an electron density of
$0.001$ a.u.\ Values are in millivolt (mV).
See caption to table \ref{table-diffs2.0} for details.
}
\label{table-diffsisorho}

\end{table}

The \DFISA method offers us a computationally efficient and numerically robust 
implementation of the ISA method. In the first paper on the ISA
method, Lillestolen and Wheatley noted that the ISA charges were remarkably 
good at describing the molecular dipole moments, often better than the 
DMA method. Of course the charges alone are not the whole story and one
needs to take into account the higher ranking multipole moments too.
This was recognised by Stone \cite{StoneA85,StoneA02,Stone05} and it is the
presence of these higher ranking terms that are the main reason for the
success of both the 1985 and 2005 DMA algorithms \cite{DayMJ05,PriceLWHPKD10}.
Consequently, a meaningful assessment of multipole moment models must include
a comparison of the higher ranking terms. But these are not unique, and this 
poses a problem: how are we to assess one model against another?
% in the absence of well-defined benchmarks? 

At distances well outside the van der Waals surface of the molecule,
we can compare the
electrostatic potential calculated from the multipole expansions ---
we will denote these as \EelstDM, the `DM' to indicate the distributed
multipole description ---
% are not expected to equal
with the reference non-expanded potential, obtained from SAPT(DFT) as the energy
\Eelst of a unit charge. The latter includes the penetration energy, which is absent in 
\EelstDM, but at large distances it will be small.

Table~\ref{table-diffs2.0} shows the mean difference,
$<\Delta> = < \Eelst - \EelstDM>$, between reference and model
electrostatic potentials at points on the surface of formamide at twice the van der Waals
radii, for various multipole models, and the standard deviation,
$\sigma_{\Delta}$, of these differences. The columns show results for models truncated at
different ranks; that is, `rank $n$' means that multipoles up to rank $n$
are included and higher multipoles discarded. For a good model, the
mean should be small but nonzero, representing the mean penetration energy,
and the standard deviation $\sigma_{\Delta}$ 
which represents the fluctuation in penetration energy over the
surface, should also be small. The values are in
millivolt; the energy of a unit charge is numerically the same value
in meV. The penetration energy here should be positive, as the
positive test charge is less screened from the nuclei as it penetrates
into the electron density.

The Table shows two variants of distributed multipole analysis.
DMA0 is the original version\cite{StoneA85} and DMA4 is
the modified version\cite{Stone05} which uses numerical integration
over a grid for terms in the gaussian expansion of the density with
exponents $\zeta$ less than 4. Finally the ISA
results were obtained using the BS-ISA+DF method described above.
%We also used the \Mulfit
%refinement\cite{WinnFR97,FerenczyWR97}, which attempts to
%improve the low-rank description by fitting high-rank multipoles on
%each atom using lower-rank ones on neighbouring atoms, but it offered
%little improvement over the standard DMA here, although it is known to be more
%successful for large molecules.

\begin{figure*}
  \includegraphics[viewport=0 30 550 340, clip, width=0.45\textwidth]{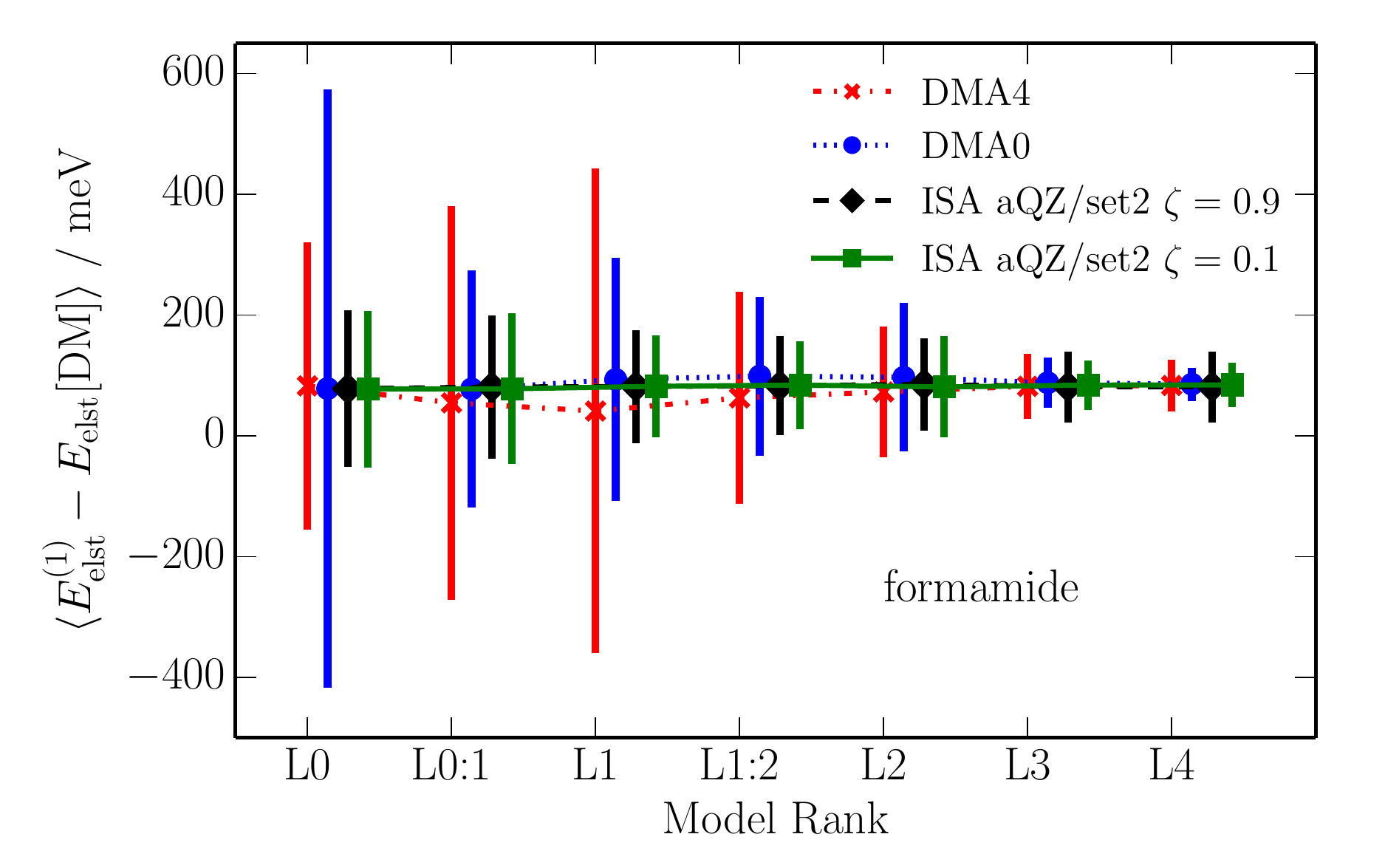}
  \includegraphics[viewport=0 30 550 340, clip, width=0.45\textwidth]{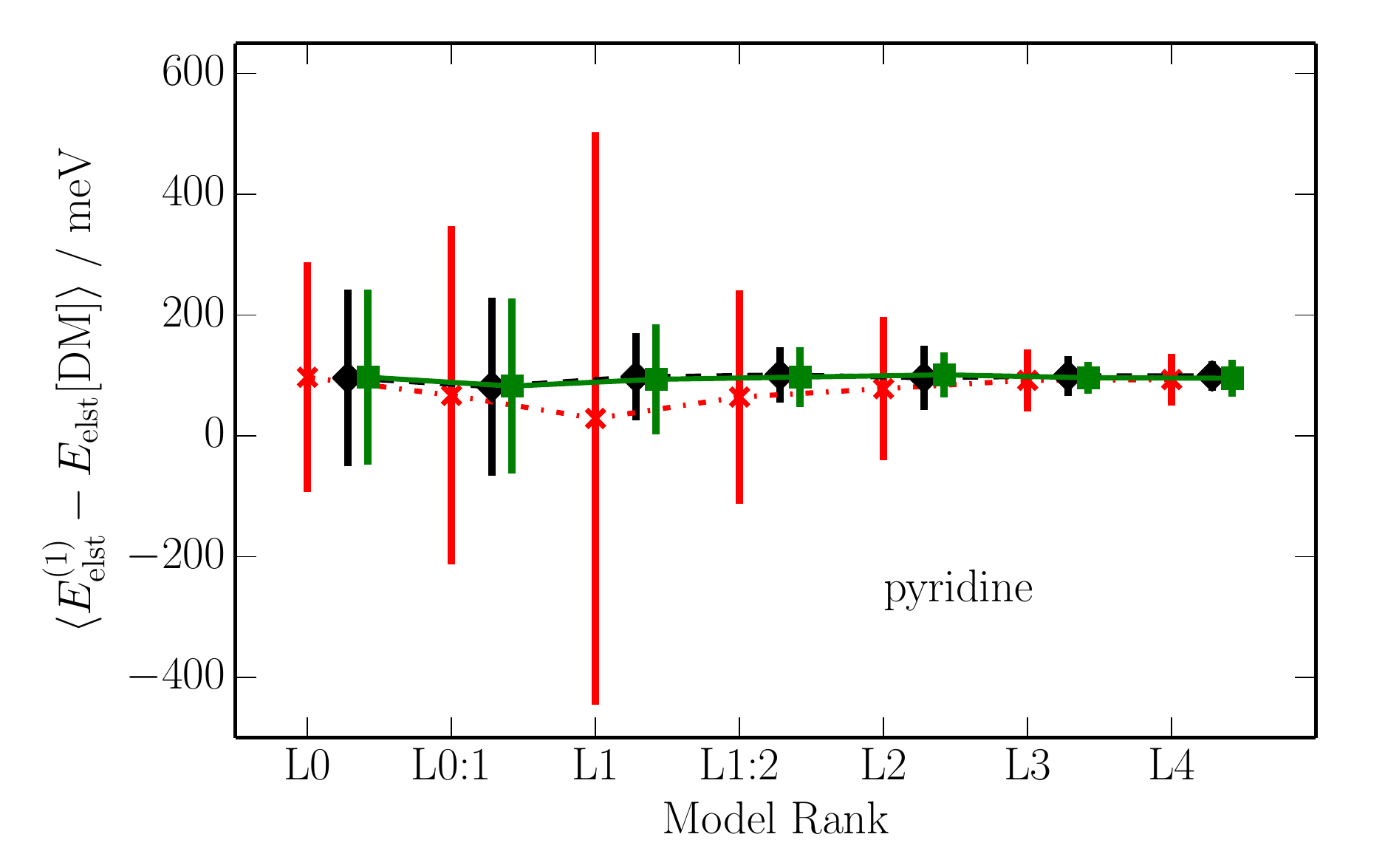}
  \includegraphics[viewport=0  0 550 340, clip, width=0.45\textwidth]{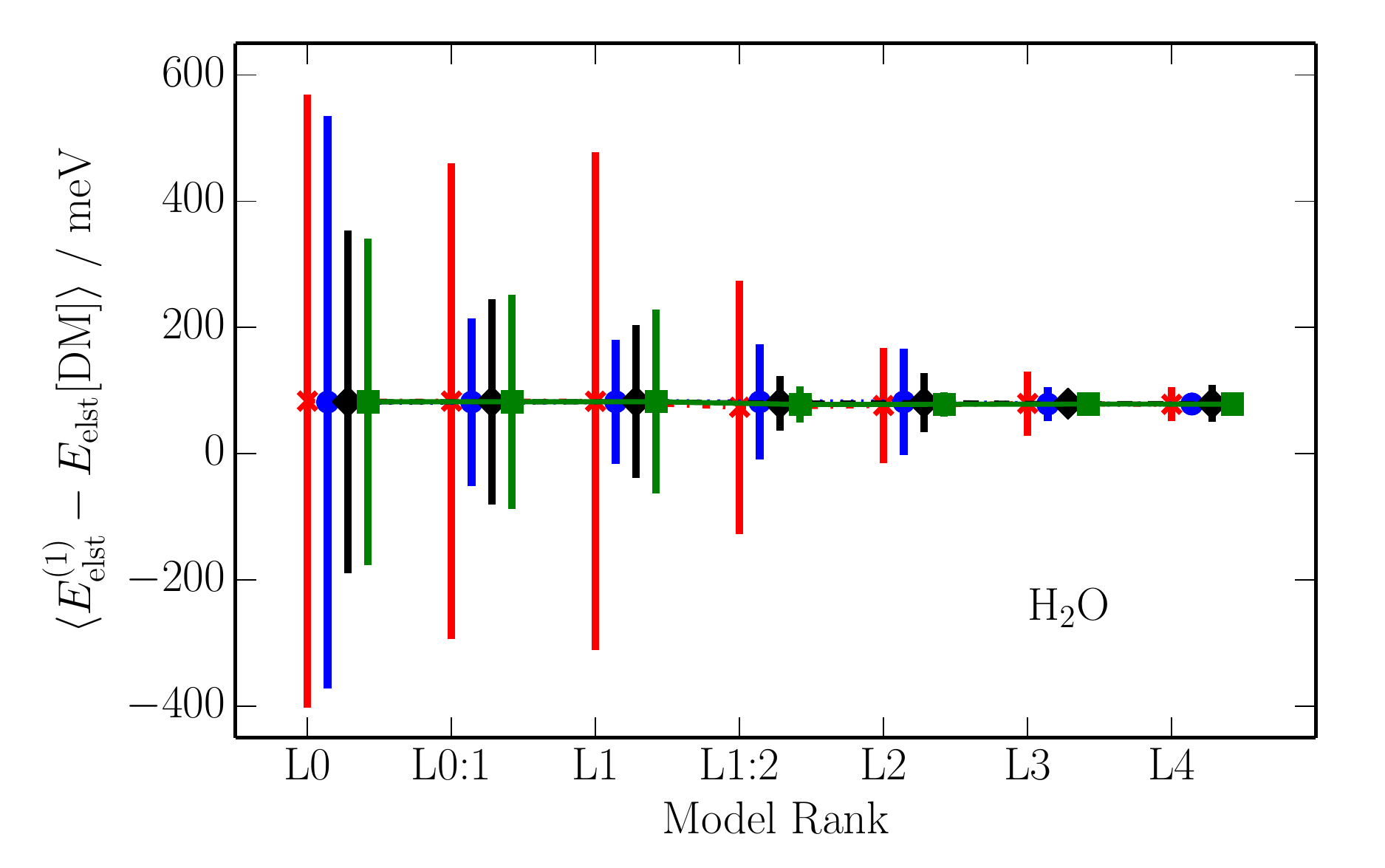}
  \includegraphics[viewport=0  0 550 340, clip, width=0.45\textwidth]{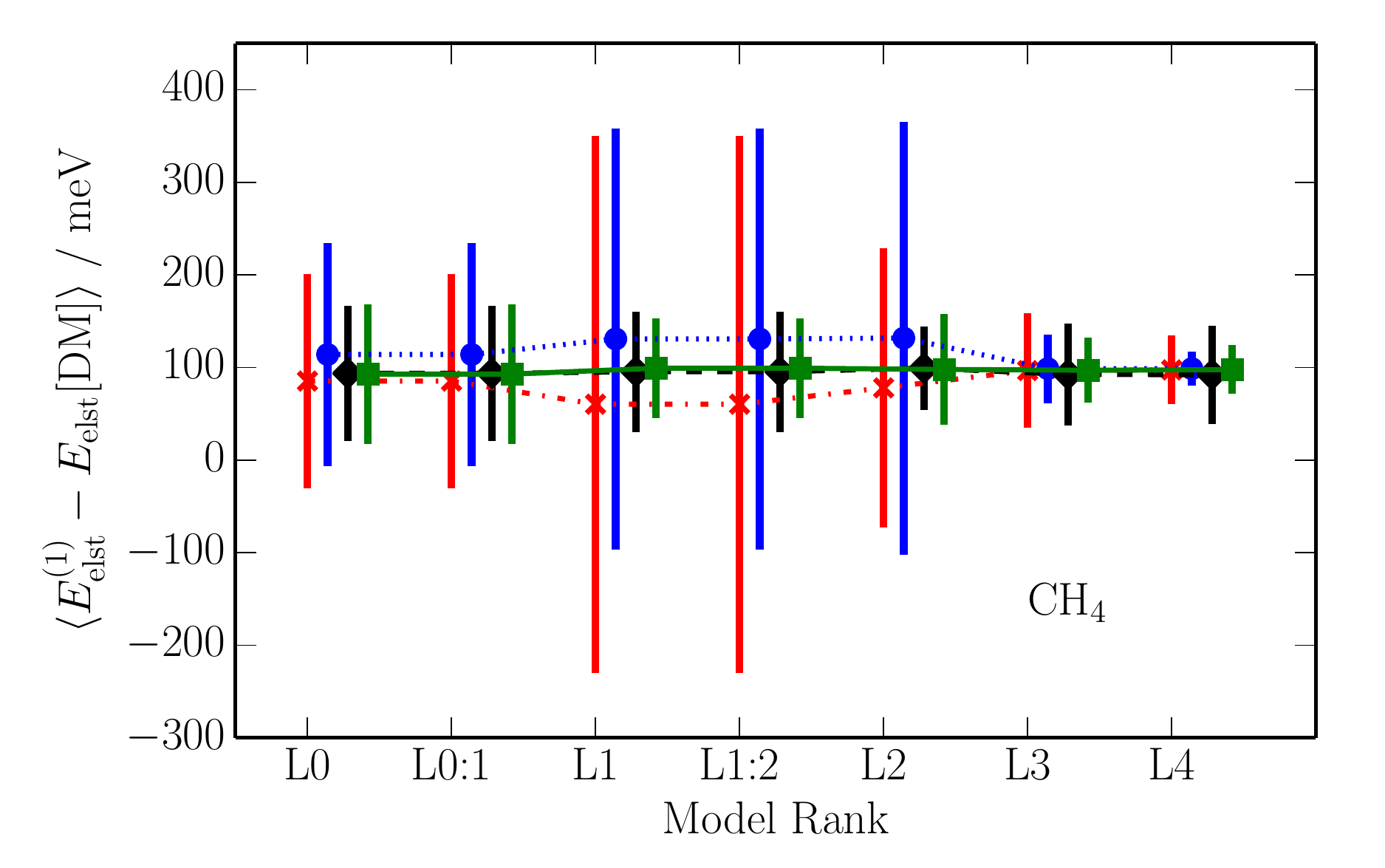}
  \caption[Energy difference statistics]{
  Average signed potential differences and and their standard deviation calculated using a point charge probe
  placed on the $10^{-3}$ iso-surfaces of the pyridine, water, formamide and methane molecules.
  The DMA0, DMA4 and \ISA $\zeta=0.1, 0.9$ models at various ranks 
  are used to calculate the multipole energies. Ranks are indicated by either 
  L$n$ or L$n:m$. In the former case, all atoms have multipoles limited to 
  rank $n$ and in the latter, hydrogen atoms are limited to rank $n$ and heavier
  atoms to rank $m$.
  \label{fig:pot_iso3_rank}
  }
\end{figure*}

Distributed multipole analysis is not recommended for use at low rank,
as local dipoles enter at rank~1, and quadrupoles, describing
$\pi$-orbital features, at rank~2, and the results at rank~0 and~1 are
poor, as expected. At rank~2 and above, however, it performs well, and the DMA0
and DMA4 variants are comparable.
It is clear, however, that the \DFISA results are substantially
better at low rank, and comparable with the DMA results at rank~2 and above.

Table \ref{table-diffs1.5} shows similar results for a surface at 1.5
times the van der Waals radii. The variation about the mean is much
greater here, but the general picture is the same. The mean
energy difference is now greater, but seems to be quite well
described by the ISA model at quite low rank.

Table \ref{table-diffsisorho} shows results on an isodensity
surface around formamide at an electron density of 0.001 a.u. Results
at rank~3 are similar to rank~4 and are not shown. The mean
differences are larger here, as expected, and they vary with rank for
DMA0 and DMA4, but are remarkably consistent for \DFISA. Moreover the
standard deviation of the differences is noticeably smaller for the
ISA method, which suggests that it may provide a promising approach
for modelling the penetration energy, and we discuss this further below.

The data are more clearly displayed in fig.~\ref{fig:pot_iso3_rank}, where
$<\Delta>$ and $\sigma_{\Delta}$ are plotted versus model.
For a multipole model with good convergence properties the average
energy difference $<\Delta>$ from the SAPT(DFT) reference
will converge quickly with rank, and the standard deviation
$\sigma_{\Delta}$ of the differences will be small.
From fig.~\ref{fig:pot_iso3_rank} it can be seen that the two DMA models 
appear to have converged only when terms up to rank 3 are included.
By contrast, with the \DFISA multipole models the average energy difference has
converged by rank 1 (charge and dipole) though $\sigma_{\Delta}$ reduces to an 
acceptable value only by rank 3, as with the DMA models. 
The \DFISA L0 model has a standard deviation $\sigma_{\Delta}$ of only $133$\,mV, about 10--15\%
more than the DMA0 and DMA4 L2 models. It would seem that, for this system, the
L0 \DFISA multipole models are competitive with the more elaborate, higher rank models.
This is also evident in fig.~\ref{fig:pot_iso3_rank} where we see that both the 
\DFISA models (with $\zeta=0.1$ and $0.9$) result in energy differences that
show very little variation with rank, though, for higher accuracies we need to 
include terms of rank 3 on to reduce the variation over the surface.

These formaldehyde data are also shown in the Supplementary Information as colour
maps of $\Delta$ plotted on the 0.001 a.u.\ isosurface of formamide.
The superior convergence pattern of the multipoles from the \DFISA 
method is clearly visible.

In considering these results, it should be borne in mind that these
figures give penetration energies for a unit charge penetrating to the
0.001 a.u. isodensity surface, which is approximately the van der
Waals surface. The corresponding energy, on these figures, is of the
order of 80 meV, or about 8 kJ/mol. In a real system, it is the
electrons of each molecule that penetrate into the other, and the
charge involved is much smaller, by a factor of at least 10, for each
atom--atom interaction, so the penetration energies will also be much smaller.

In fig.~\ref{fig:pot_iso3_rank} we display similar data for three other systems:
water, pyridine and methane. The DMA0 models are not included for pyridine as
the energy differences obtained with this model are too large to be meaningfully
displayed along with energies from the other models. The broad conclusions
reached with the formamide system are seen to hold for all molecules: 
the \DFISA models (with $\zeta=0.1$ and $0.9$) result in energy differences
which exhibit the fastest convergence with rank and the smallest
variation over the surface.
On the other hand, the DMA0 and DMA4 models show considerably more erratic 
average energy differences and significantly larger values of $\sigma_{\Delta}$.
However, the L3 and L4 models from all four methods tend to be reasonably close, 
with similar values of $<\Delta>$ and $\sigma_{\Delta}$. 

\subsection{Assessing the models using molecular dimers}
\label{sec:assessment_using_dimers}

From the discussion above it should be clear that the \DFISA multipole models
show better convergence behaviour than the DMA models.
It is also evident from the data presented in \ref{fig:pot_iso3_rank}
that the \DFISA point charge (L0) models are better than those from both DMA 
algorithms. This is not unexpected, as the ISA algorithm 
guarantees the most spherical atomic domains (within precision and 
algorithmic implementation), and the DMA method does not claim to
produce useful charge models. However a pertinent question is whether
these L0 models with charges on atomic sites only can be used for modelling
the electrostatic interaction, and what kinds of error should we expect if this is
done. This issue is particularly pertinent as many simulation programs
are not able to use anything other than point charges, and, in any case,
for large simulations involving biologically interesting molecules, it is 
often not feasible to use higher ranking multipoles due to the computational
cost incurred. To fully address the questions associated with point charge
models would take us too far from the central aim of this paper, so 
we just outline the issues.
Further, from fig.\ref{fig:pot_iso3_rank} it may seem that the DMA0, DMA4 and two \DFISA models 
are nearly equivalent when high ranking multipoles are included: the L3 and L4 
models from these methods appear very similar. However these results were
obtained using a point-charge probe interacting with the molecule and some
differences between the models are not picked up in this way. 

\begin{figure*}
  \includegraphics[viewport=0 37 550 340, clip, width=0.45\textwidth]{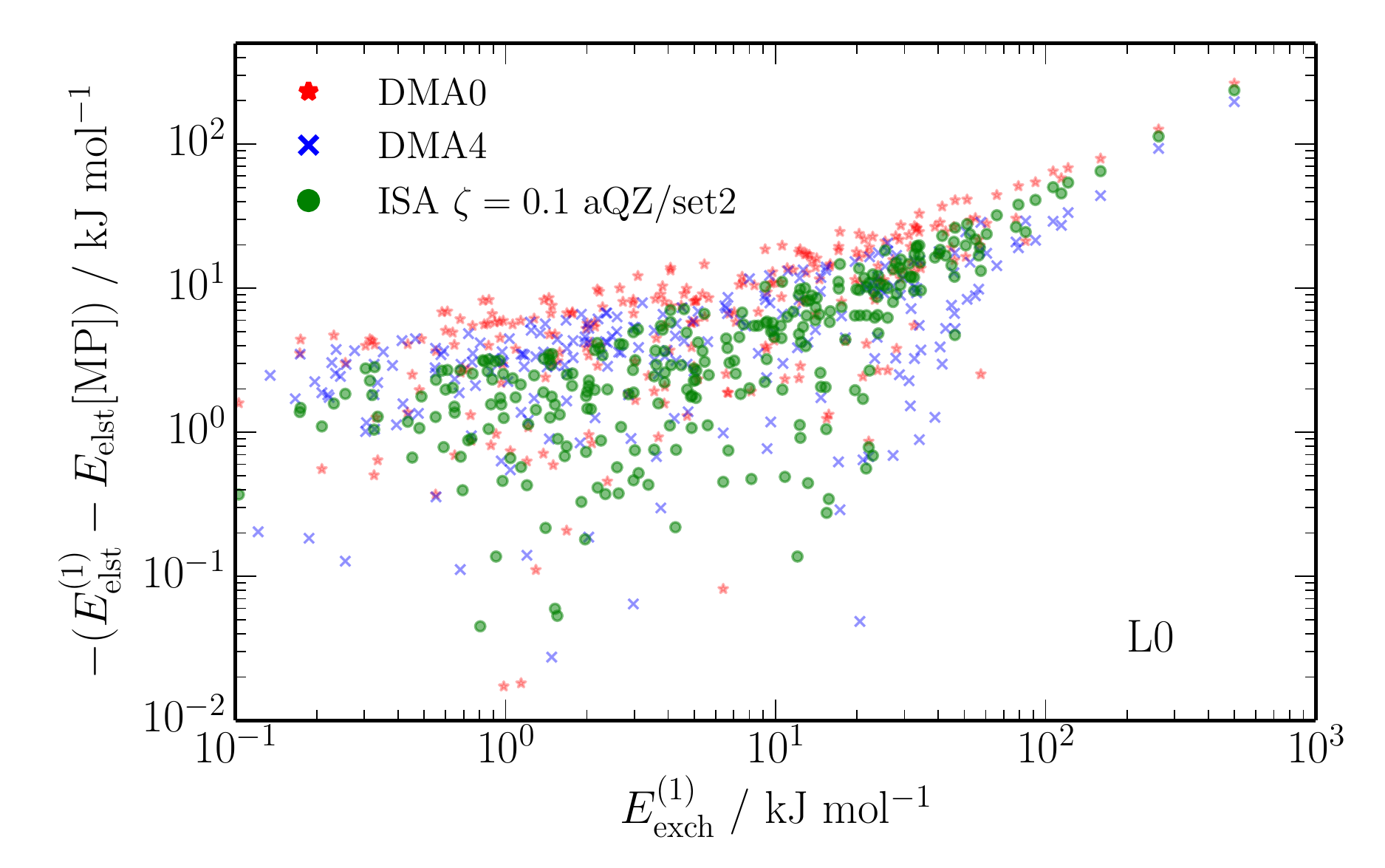}
  \includegraphics[viewport=0 37 550 340, clip, width=0.45\textwidth]{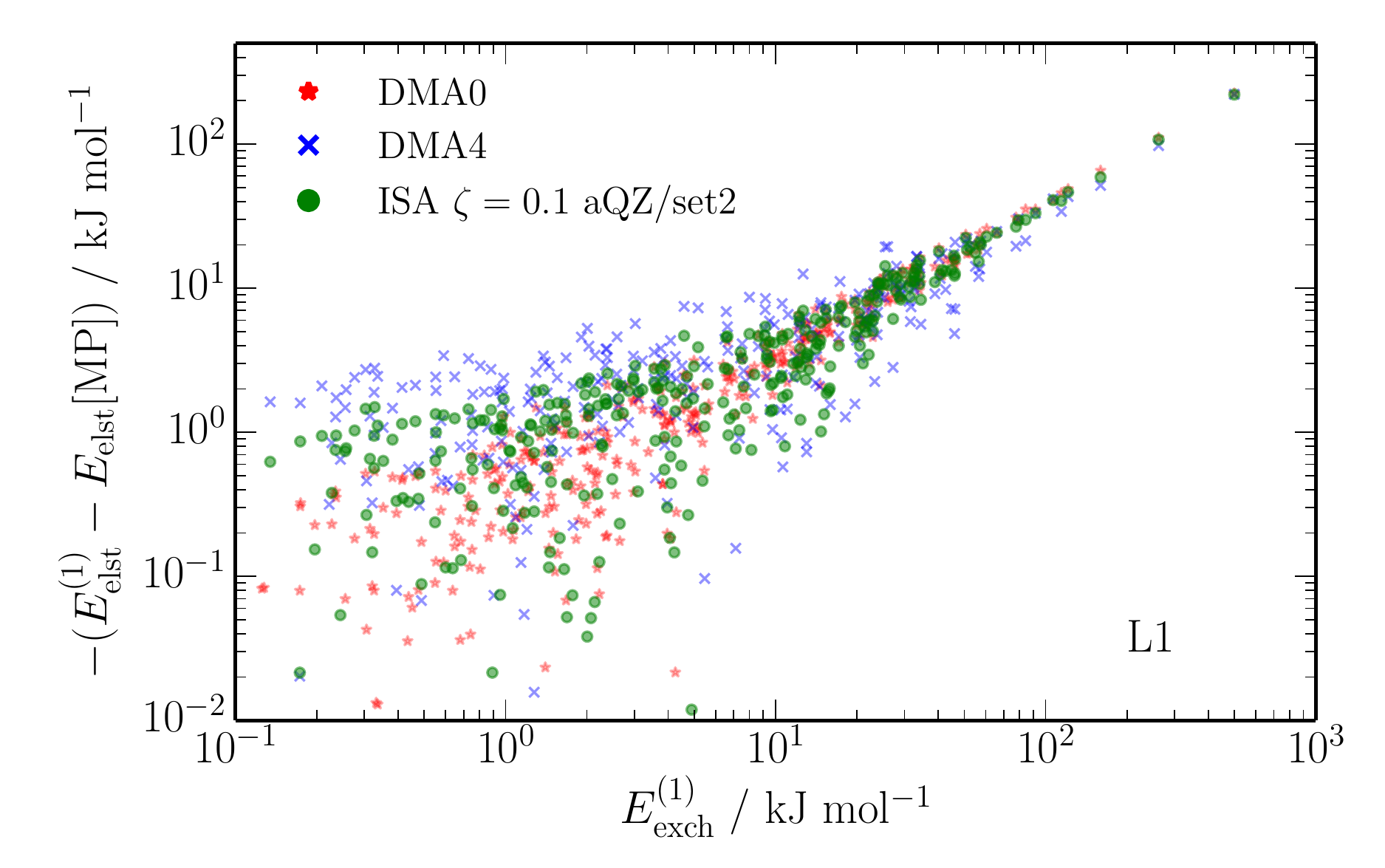}
  \includegraphics[viewport=0 10 550 340, clip, width=0.45\textwidth]{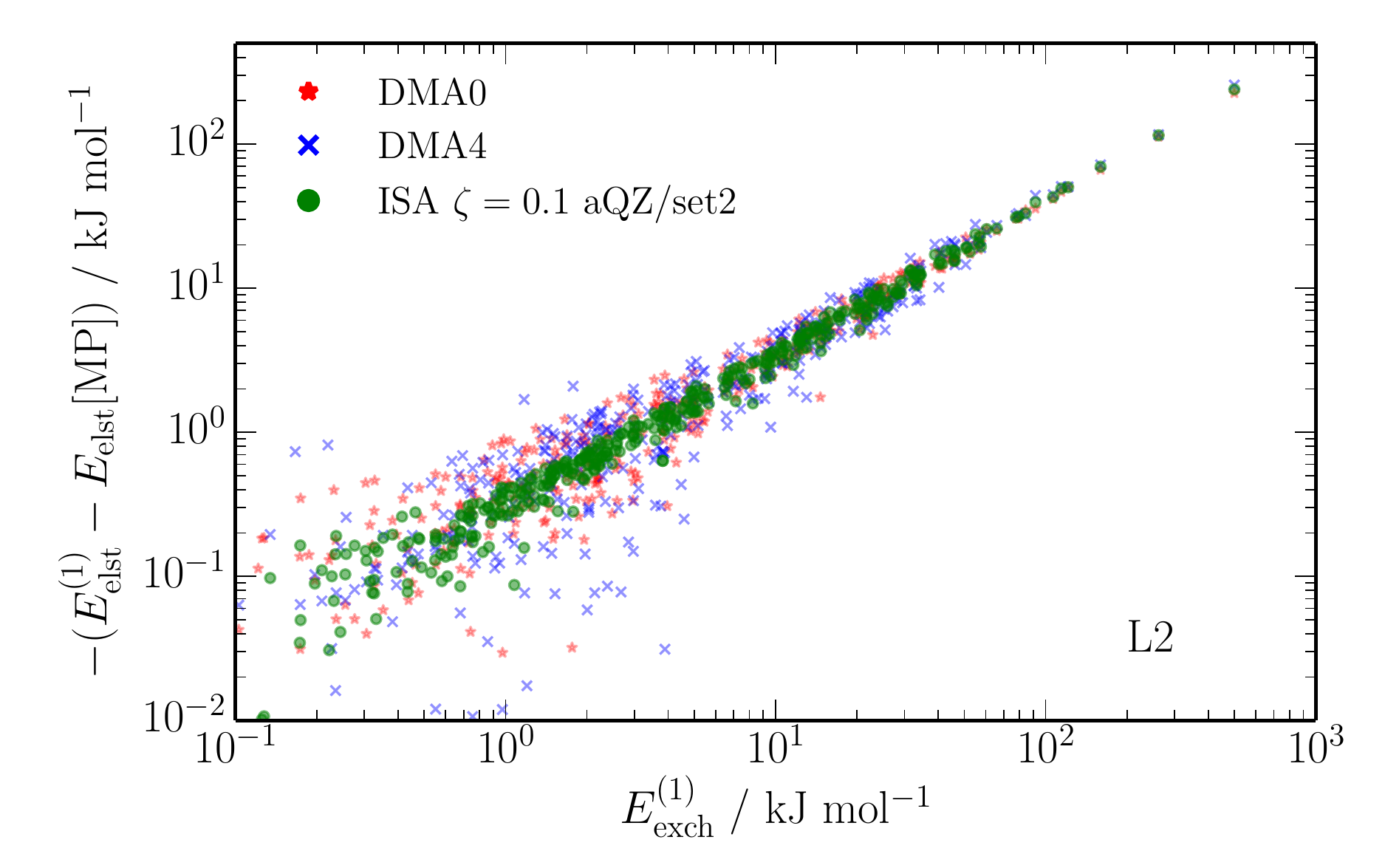}
  \includegraphics[viewport=0 10 550 340, clip, width=0.45\textwidth]{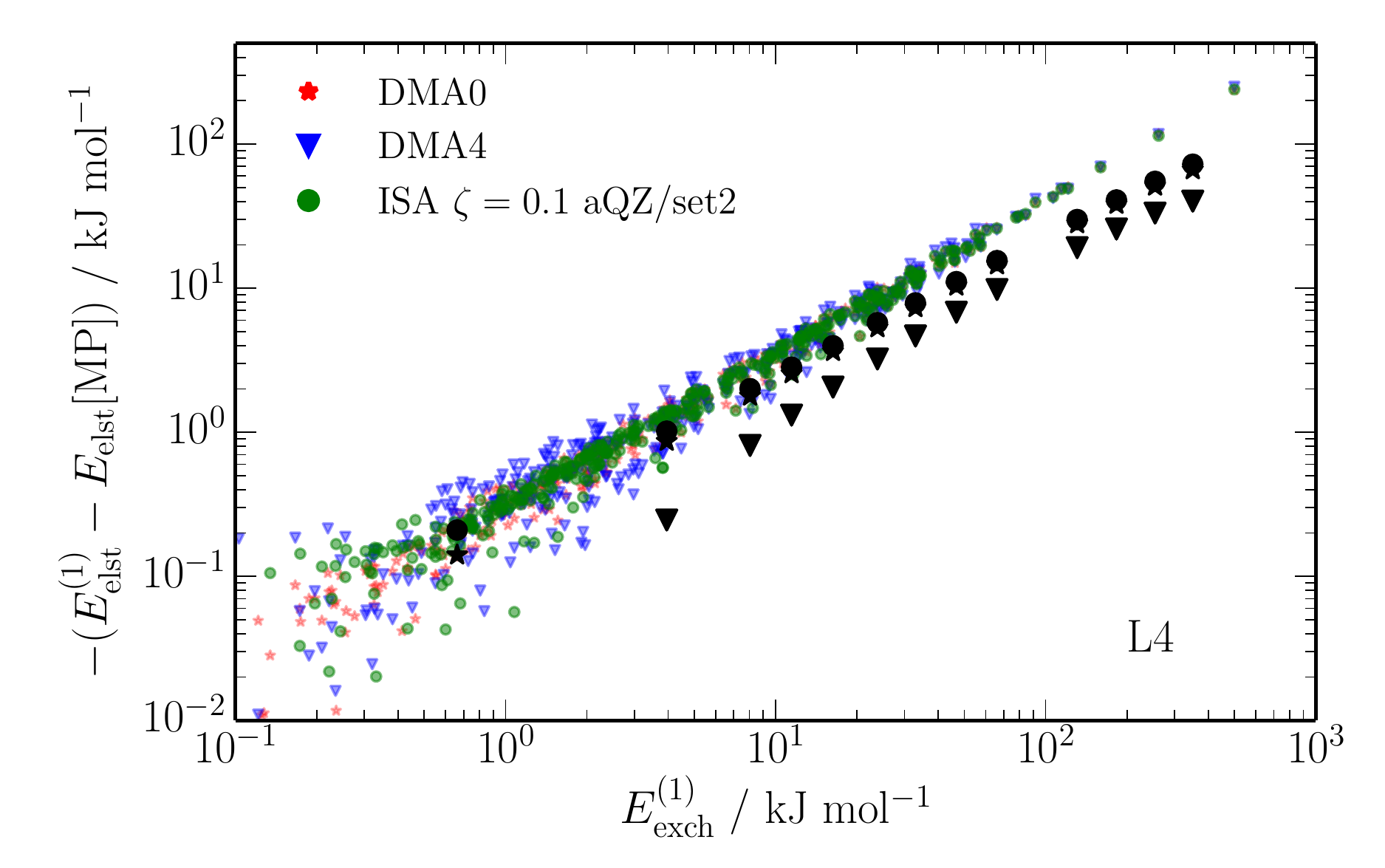}
  \caption[Water dimer multipole errors]{
  Multipole errors for the water dimer in 400 pseudo-random dimer geometries.
  The multipole error is plotted against the first-order exchange-repulsion \Eexch.
  The multipole models have been calculated using the aug-cc-pVQZ main basis and the 
  SAPT(DFT) \Eelst and \Eexch energies have been calculated using the
  aug-cc-pVTZ basis in the MC+ type. See the text for details.
  In the rank 4 panel we have additionally included data (large black points)
  for the water dimer in its minimum energy dimer orientation.
  \label{fig:water2_pen_random}
  }
\end{figure*}

\begin{figure*}
  \includegraphics[viewport=0 37 550 340, clip, width=0.45\textwidth]{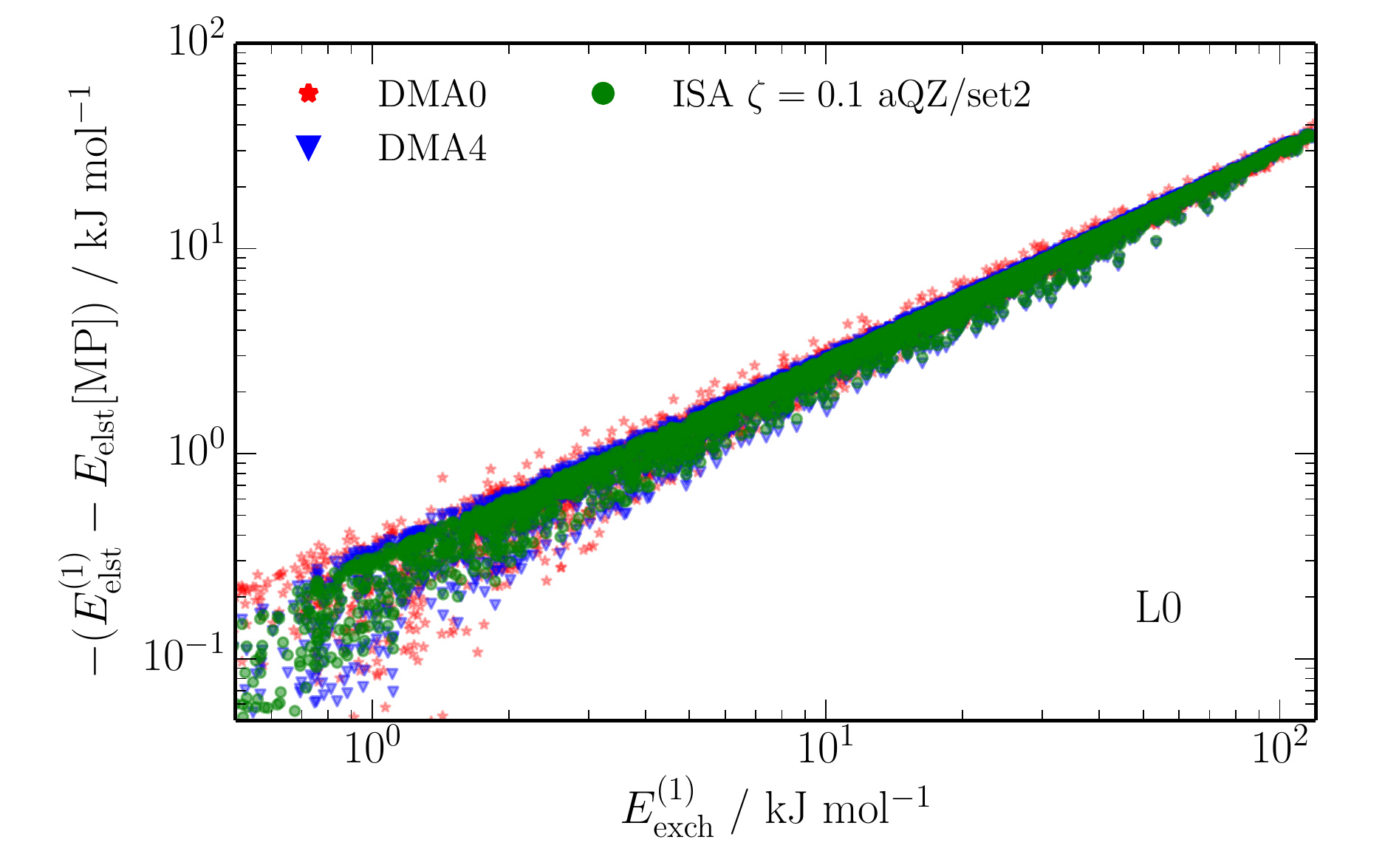}
  \includegraphics[viewport=0 37 550 340, clip, width=0.45\textwidth]{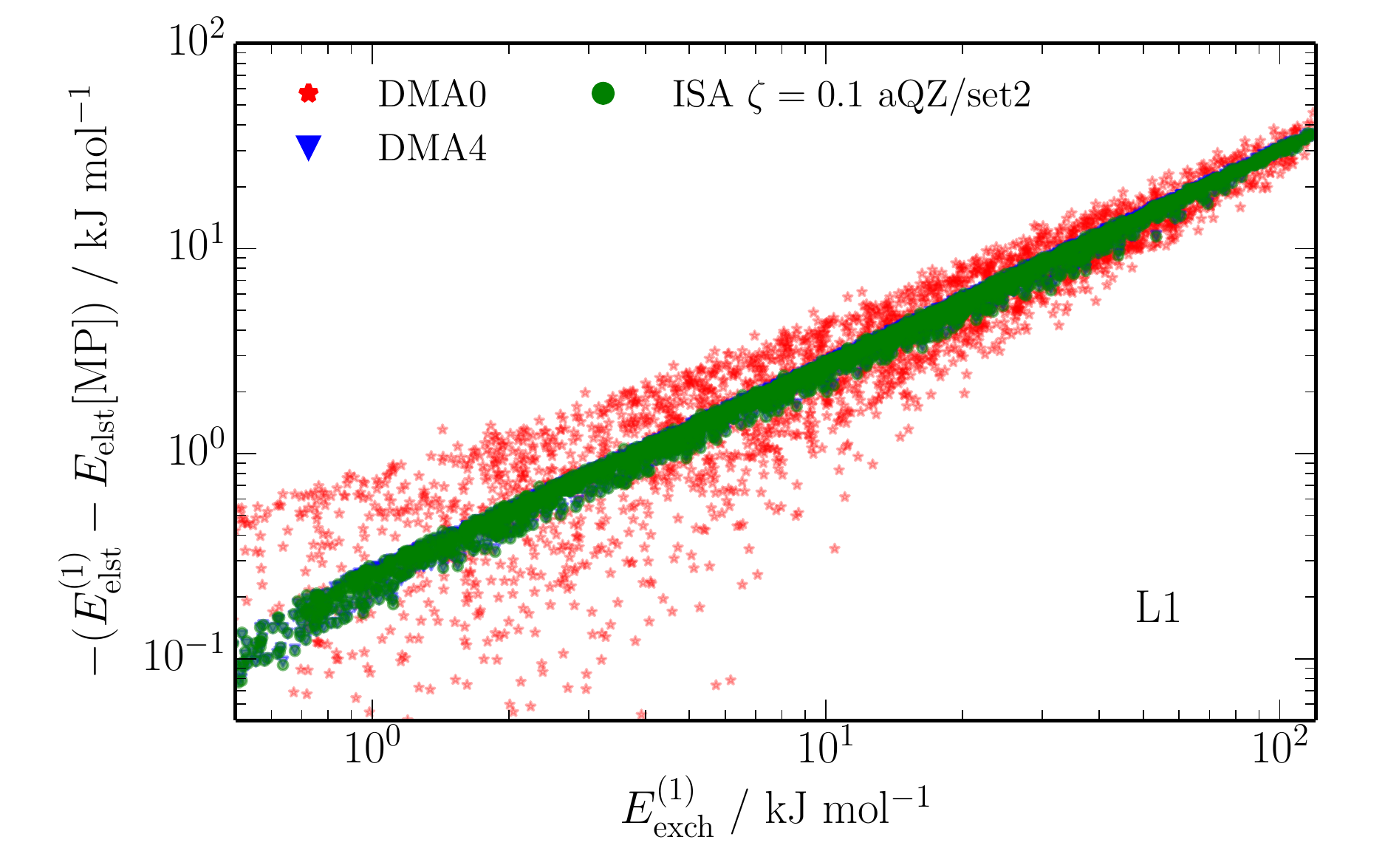}
  \includegraphics[viewport=0 10 550 340, clip, width=0.45\textwidth]{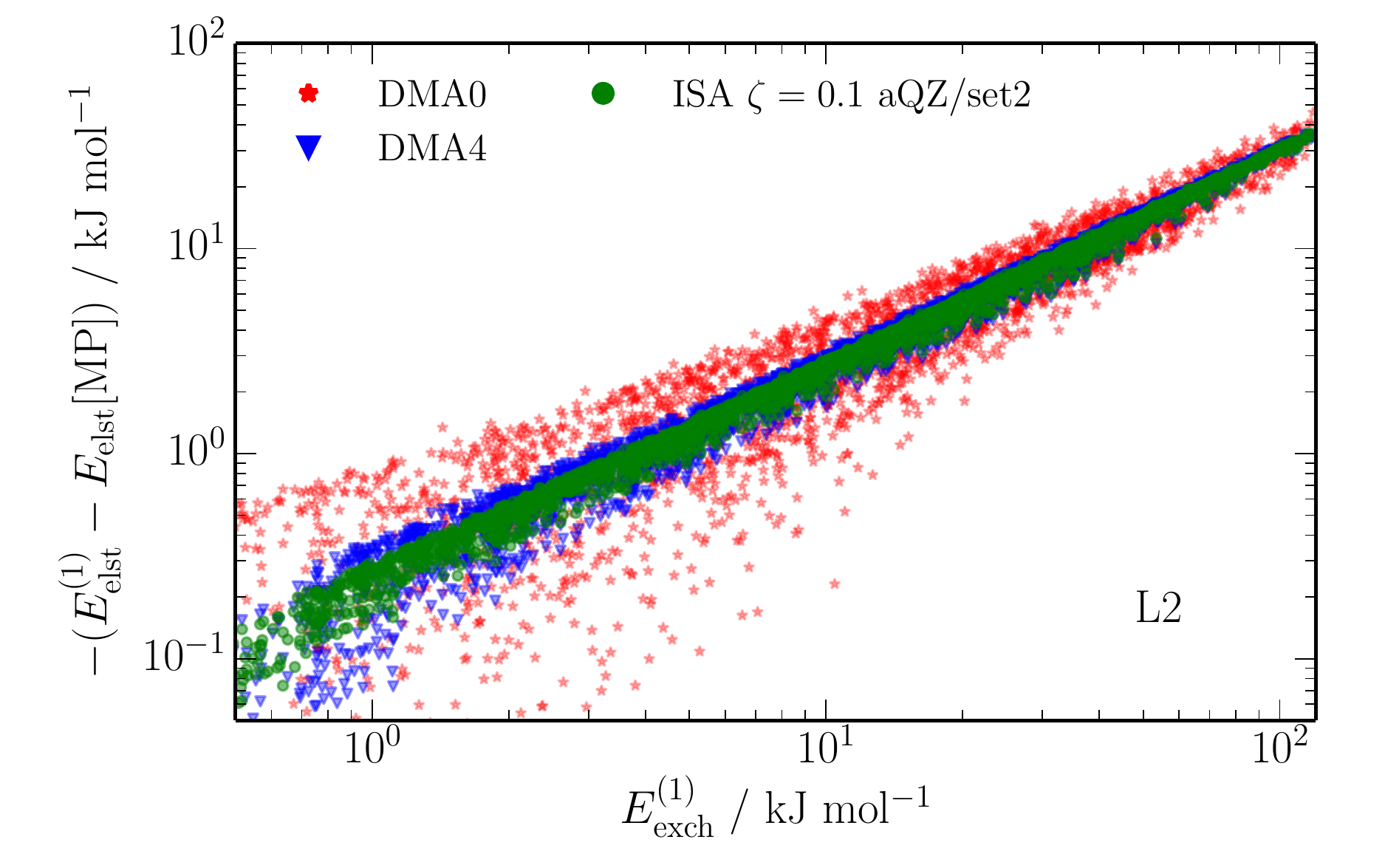}
  \includegraphics[viewport=0 10 550 340, clip, width=0.45\textwidth]{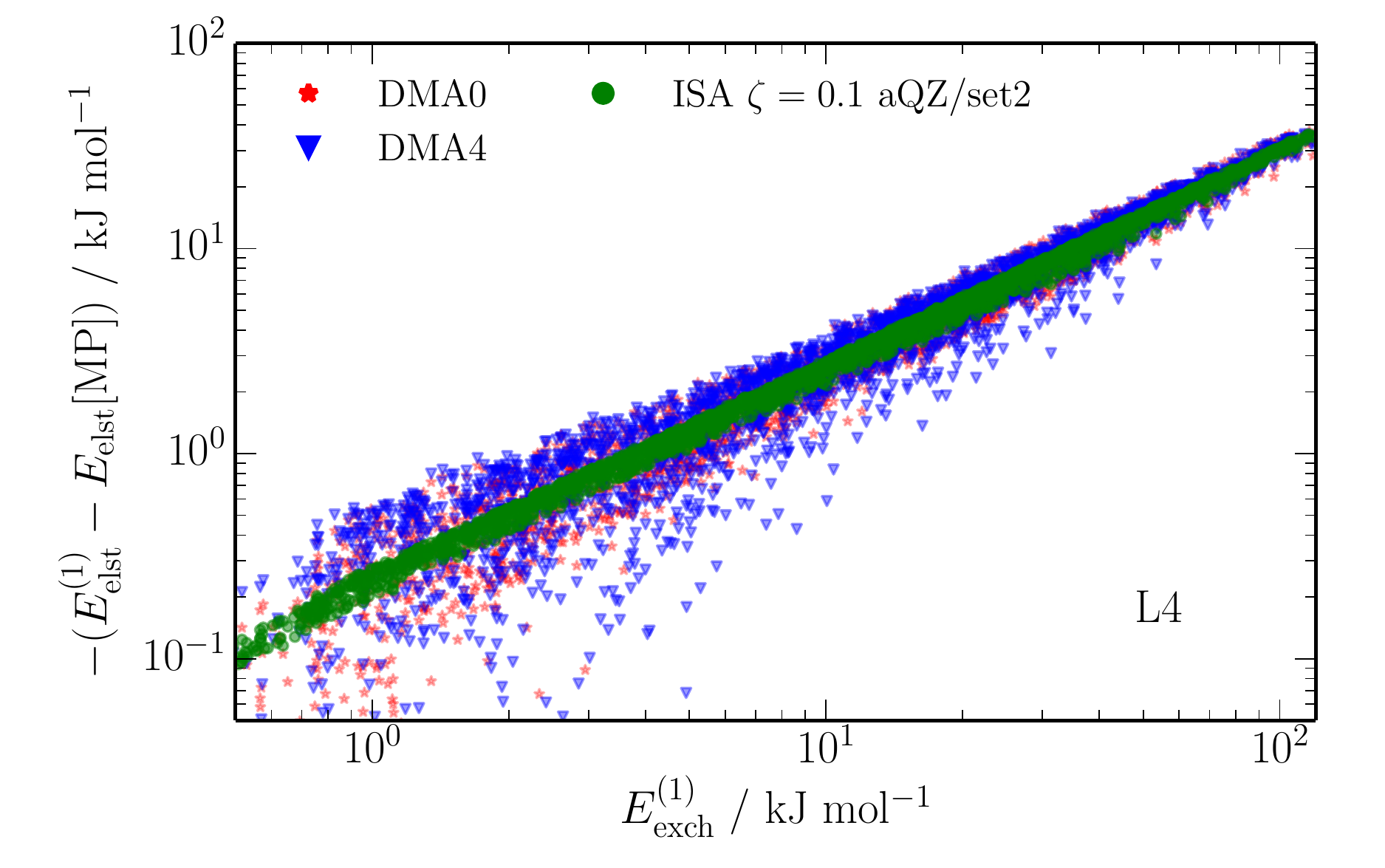}
  \caption[Methane dimer multipole errors]{
  Multipole errors for the methane dimer in 2600 pseudo-random dimer geometries.
  The multipole error is plotted against the first-order exchange-repulsion \Eexch.
  The multipole models have been calculated using the aug-cc-pVQZ main basis and the 
  SAPT(DFT) \Eelst and \Eexch energies have been calculated using the
  aug-cc-pVTZ basis in the MC+ type. See the text for details.
  \label{fig:methane2_pen_random}
  }
\end{figure*}

\begin{figure*}
  \includegraphics[viewport=0 37 550 340, clip, width=0.45\textwidth]{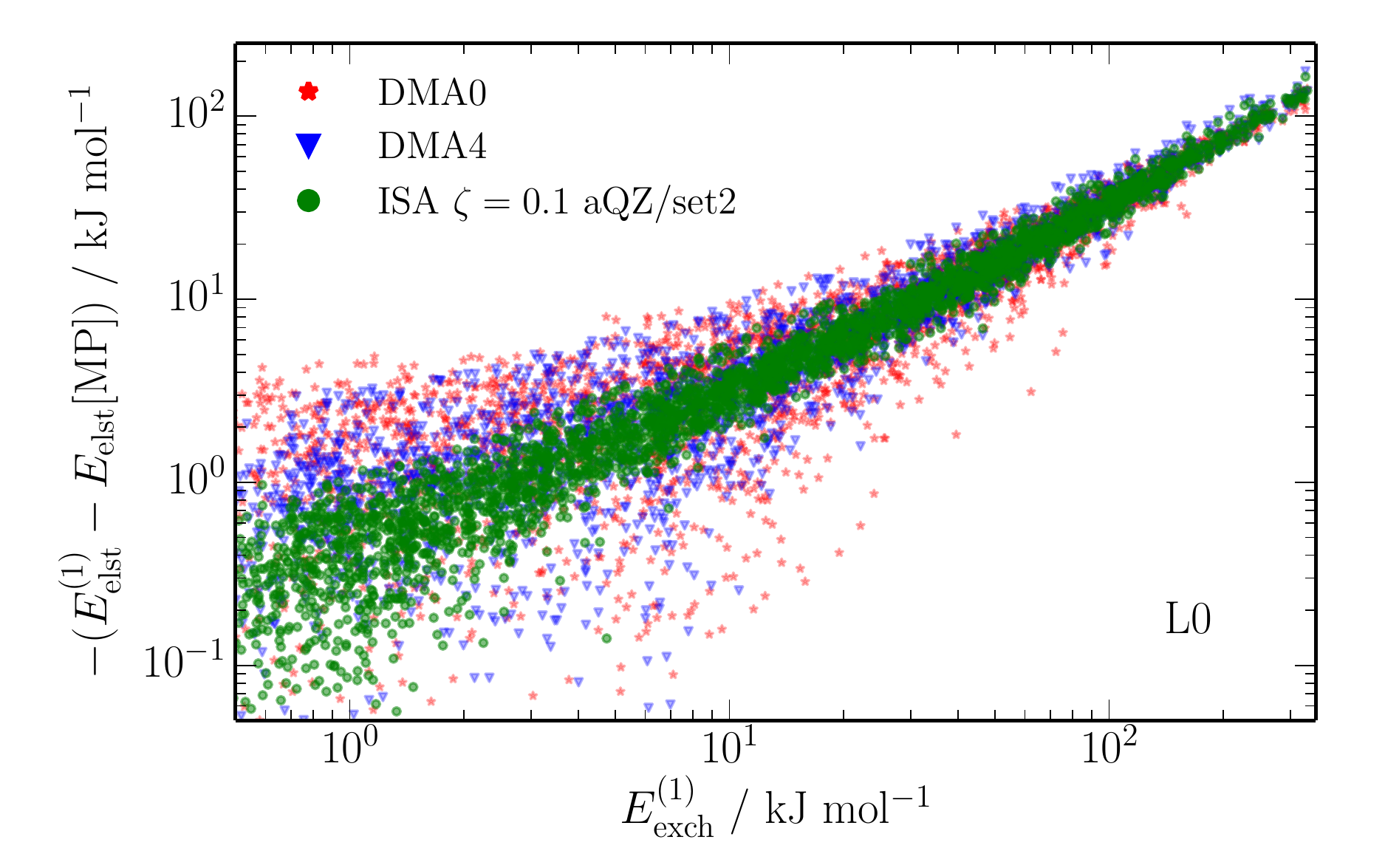}
  \includegraphics[viewport=0 37 550 340, clip, width=0.45\textwidth]{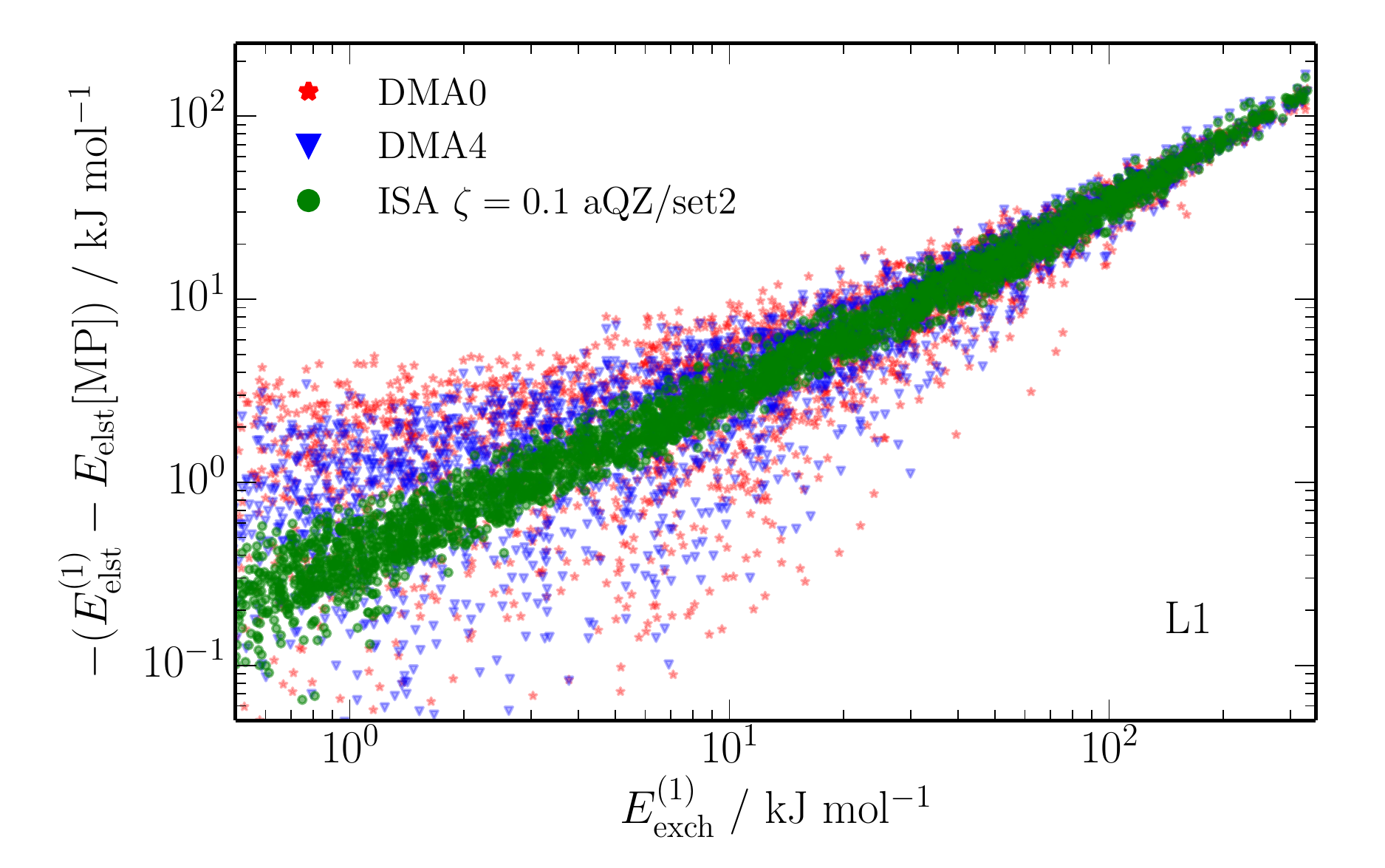}
  \includegraphics[viewport=0 10 550 340, clip, width=0.45\textwidth]{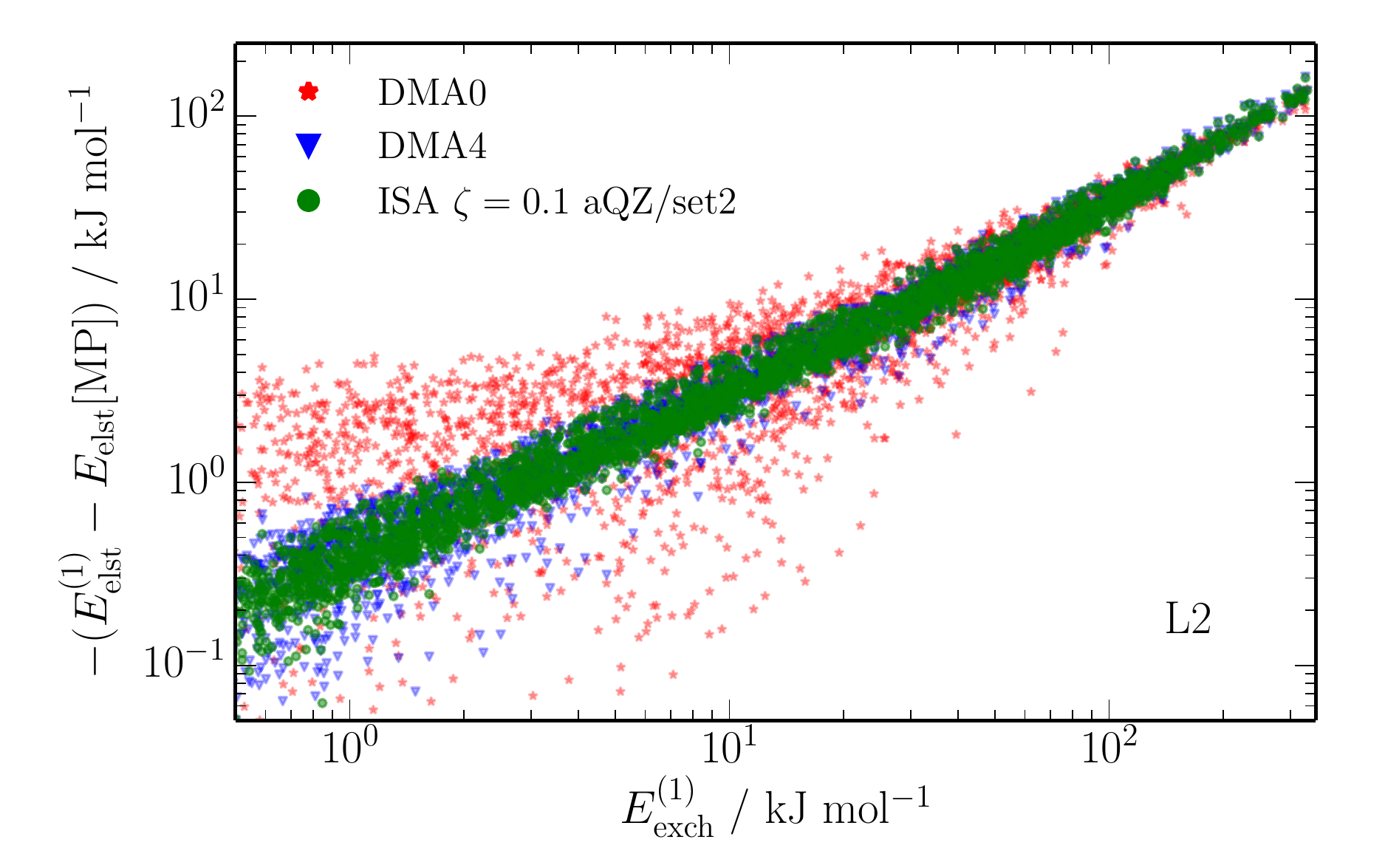}
  \includegraphics[viewport=0 10 550 340, clip, width=0.45\textwidth]{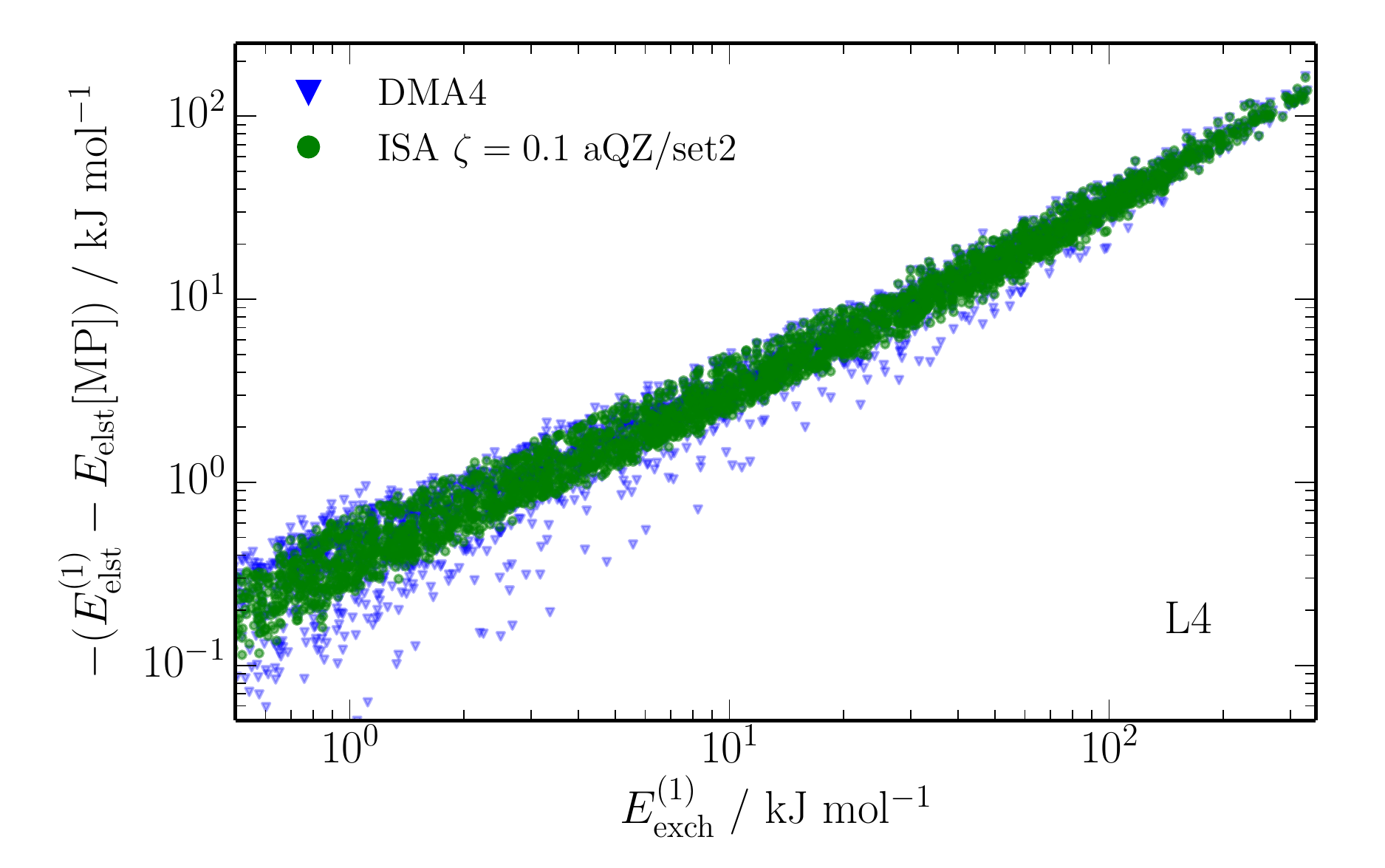}
  \caption[pyridine dimer multipole errors]{
  Multipole errors for the pyridine dimer in pseudo-random dimer geometries.
  The multipole error is plotted against the first-order exchange-repulsion \Eexch.
  The multipole models have been calculated using the aug-cc-pVQZ main basis and the 
  SAPT(DFT) \Eelst and \Eexch energies have been calculated using the
  Sadlej pVTZ basis in the MC type.
  The scatter in the DMA0 penetration energies does not change with rank, consequently
  these results are not shown in the rank 4 panel so as to better highlight the DMA4 
  penetration energies.
  \label{fig:pyr2_pen_random}
  }
\end{figure*}

We now address these issues using energies calculated for the water, pyridine and
methane dimers at a variety of orientations.
In figs.~\ref{fig:water2_pen_random}, \ref{fig:methane2_pen_random} and 
\ref{fig:pyr2_pen_random} we display, for the DMA0, DMA4 and \DFISA($\zeta=0.1$)
multipole models at various ranks, the difference between the SAPT(DFT)
electrostatic energy and energies calculated using the models.
For brevity, we refer to this difference as the `multipole error'.
It includes deficiencies in the multipole model as well as
the penetration energy. Since it is usually assumed that the
penetration energy is proportional to the first-order exchange energy, 
the energy differences are plotted against the first-order exchange energy
\Eexch, and we expect a straight line if the assumption is correct and if
the multipole model is good enough.
Plotting the results in this way also illustrates the smaller energies
and the differences between the models more clearly.
Note that, for convenience we plot $-\Delta = - (\Eelst - \EelstDM)$
against \Eexch in these figures.

First of all consider the data for the water dimer. 
In fig.~\ref{fig:water2_pen_random} we plot $\Delta$ against \Eexch for 400 
pseudo-random water dimer configurations generated using the \CamCASP program
\cite{MisquittaWSP08,CamCASP}.
From fig.~\ref{fig:water2_pen_random} 
we see that all three L0 (point charge) models show a very poor correlation between 
$\Delta$ and \Eexch. This should not be a surprise as a considerable body of work has shown
that to model the electrostatic energy accurately in small molecules like water,
additional charges are needed on off-atomic sites. For example, most 
accurate models of water use five sites for the electrostatic model (see for example
the SAPT-5s potential\cite{MasBSG00}).
The correlation between $\Delta$ and \Eexch improves only when we include terms
to rank 2 on all sites.
At all ranks, the \DFISA multipoles result in the best correlation of $\Delta$ and \Eexch
with the correlation growing as rank increases. 
At rank 4 the DMA0 and \DFISA models result in similar energies, but the DMA4 model
exhibits somewhat more scatter.
Since $\Delta$ approaches the penetration energy as the 
rank of the model increases, we may state that there is an approximate correlation
between the penetration energy and the first-order exchange energy. This 
has been observed before, but there does not appear to be a proof of this correlation.

The relationship between the penetration energy (here taken to be the L4 $\Delta$ 
energies) and \Eexch appears to be linear for a {\em fixed dimer orientation}, 
but the constant of proportionality is dependent on the dimer geometry.
The orientational dependence of this proportionality is illustrated in the
last panel (at rank 4) in fig.~\ref{fig:water2_pen_random} where we include data for
dimers in the hydrogen-bonded configuration of water. 
The \DFISA and DMA0 models both result in similar $\Delta$ energies that show a linear
correlation with \Eexch, but with a different constant of proportionality compared
with the high-energy configurations from the pseudo-random data set.
The DMA4 model once again yields the worst correlation and may not be 
suitable for the water molecule.

A consequence of the above observations is that we should expect the best correlation
between the penetration energy and \Eexch for systems that are close to
spherical, such as the methane molecule. In fig.~\ref{fig:methane2_pen_random}
we plot $\Delta$ against \Eexch for the methane dimer in 2600 pseudo-random 
configurations. 
Two features stand out: the \DFISA multipole model results in a 
strong correlation between $\Delta$ and \Eexch at all ranks, but by
rank 4 (rank 3 results are similar) the correlation is nearly perfect.
On the other hand, while the DMA0 and DMA4 models are comparable to the
\DFISA model at rank 0, there is no systematic behaviour of these
models at higher ranks: the L1 DMA4 and \DFISA models are nearly identical
but when terms beyond rank 2 are included, $\Delta$ from both DMA models 
is poorly correlated with \Eexch.
We note here that it is possible to improve the quality of the DMA4 multipoles
by reducing the value of the Becke smoothening parameter \cite{Stone05}  used to perform
the real-space partitioning of the most diffuse functions, but it is not
clear if this strategy can be expected to work more generally for other systems.

Finally, in fig.~\ref{fig:pyr2_pen_random} we plot similar data for the 
pyridine dimer in around 3000 orientations. Pyridine is a highly anisotropic
system so we should expect a strong orientational dependence in the proportionality
between $\Delta$ and \Eexch. This does seem to be the case. The correlation 
between these two energies is not as good as either the water dimer or the 
methane dimer, nevertheless, here too, it is the \DFISA model that yields the
best correlation between $\Delta$ and \Eexch at all ranks. The DMA0 multipoles
are poor at all ranks and the DMA4 model is a considerable improvement, though
even this model cannot compete with the \DFISA, even when terms to rank 4 are included.
Notice that once again the \DFISA model exhibits the fastest convergence with
rank: the charge only model may be adequate for many purposes and we see
nearly converged results when terms of rank 1 (dipoles) are included. 
The \DFISA model is essentially fully converged by rank 2. These observations
are in-line with those made from the data plotted in fig.~\ref{fig:pot_iso3_rank}.

\section{Analysis}
\label{sec:analysis}

We have described and presented results from a numerically stable and robust
implementation of the iterated stockholder atoms (ISA) approach of 
Lillestolen and Wheatley \cite{LillestolenW08}. This approach, termed the 
\DFISA method, works entirely in basis-space and can be combined with 
standard density-fitting functionals using a single parameter $\zeta$ that
controls the relative weights of the \ISA and density-fitting functionals.

The \DFISA method uses auxiliary basis sets that are substantially larger that
those normally used for density-fitting. In particular, the s-functions
sets which are needed to define the ISA shape functions are unusually
over-complete and flexible. This is needed as a considerable degree of
variational flexibility is required to minimize the ISA part of the 
\DFISA functional. We have demonstrated that with smaller, more
inflexible basis sets, the functional minimum does not correspond to the
true minimum of the ISA functional. In particular, the shape functions are
not well defined in the tail region.

The \DFISA functional is shown to converge in less than 80 iterations and
sometimes as few as 10. Convergence is exponential with iteration 
number and seems to be independent of the molecular size or type of basis
used.
In contrast, conventional methods for solving the ISA equations either
work in real-space and converge (if at all) in 1000 iterations or so,
or partially work in an excessively restricted basis-space and converge to a 
false minimum in 140 iterations or so.

The numerical implementation of the \DFISA functional is identical with 
that of conventional density-fitting functionals, so it can easily be applied
to systems of hundreds of atoms. Additionally, the pure ISA part of the
\DFISA functional scales linearly with the number of atoms.
This feature, and the overall high accuracy and good convergence
properties of the \DFISA functional should make it ideally suited for 
applications to large molecules. One restriction, though, is that 
since we rely on Gaussian (finite-extent) basis sets, this functional 
cannot as yet be used with infinite systems.

The main goal of this paper has been to investigate the applicability of the
ISA method as an alternative for distributed multipoles. Having a stable
ISA implementation was essential for this as both the low and high ranking 
distributed multipoles are sensitive to the partitioning method, particularly
to the way in which the atom-like density tails are modelled. 
This motivation was central to the attention we have paid to converging
the atomic density tails in the \DFISA functional.

We have used the \DFISA method to calculate distributed multipoles
and have compared electrostatic energies and multipole errors
computed with these multipoles and those from
distributed multipole analysis; both the 1985 version
that works entirely in basis-space \cite{StoneA85} giving the 
DMA0 multipoles, and the 2005 version that works partially 
in real-space and has better stability with basis sets
\cite{Stone05} giving the DMA4 multipoles. 
For the dozen systems we have studied the \DFISA multipoles are
found to give uniformly better results at low rank than both the DMA0 and DMA4 models.
That is, if multipoles are limited to low rank (below~3), the \DFISA models
are found to result in better-converged electrostatic/penetration 
energies than either the DMA0 or DMA4 models. Further, while the 
best \DFISA results are obtained with $\zeta=0.1$, that is, 10\% ISA and
90\% density-fitting functionals, the variation with $\zeta$ is 
usually insignificant and decreases with an increase in the variational
flexibility of the auxiliary basis sets used in the minimization of the
\DFISA functional.

At rank 0 (charges only) the \DFISA models are the most accurate we
have obtained. They are often substantially more accurate than
the DMA models at rank~0, but it must be emphasized that the DMA
method is explicitly not intended for developing point charge models.
The \DFISA models have been shown to exhibit the fastest convergence
with rank of the multipole expansion. Further, in contrast to the
DMA0 and DMA4 models, the \DFISA multipoles converge systematically
with increasing rank. Therefore, these models can be truncated to lower
ranks without erratic increases in the errors incurred.
Additionally, the penetration energy---defined as the difference in the
non-expanded electrostatic energy \Eelst and the multipole energy
calculated using converged \DFISA model---exhibits a uniformly
excellent correlation with the first-order exchange energy \Eexch. 
This property makes the \DFISA multipole models an excellent choice for
building intermolecular potentials, as the penetration energy is 
often fitted together with the first-order exchange \cite{StoneM07,MisquittaWSP08}.

The numerical superiority of the \DFISA distributed multipoles makes them
potential replacements for the DMA0 and DMA4 multipoles, which 
have set the benchmark for accuracy for the last thirty years. 
This has been a high benchmark to surpass but there were already 
indications from Lillestolen and Wheatley's original paper that the ISA 
might surpass the DMA and we find that this appears to be the case.
However the original DMA method has some advantages over both 
its successor, DMA4, and the \DFISA method: it is numerically exact and 
computationally simple and it is very fast, even for large molecules.
Furthermore, both DMA methods allow 
the addition of off-atomic expansion centres, which are not possible 
in the ISA approach. These may be reasons enough to use the DMA methods for
certain applications, but from the evidence provided here we suggest that
the ISA, particularly in the \DFISA implementation shown here, is better
suited for highly accurate, rapidly convergent, and perhaps, even more
physically appealing, distributed multipole expansions.

\section{Additional information}
All developments have been implemented in a developer's version of the
\CamCASP 5.8 \cite{CamCASP} program which may be obtained from the
authors on request. 
The supplementary information (SI) contains additional data from 
the systems we have investigated but not included in this paper.
Additional information about the basis sets used is also included in the
SI.

\section{Acknowledgements}
AJM and FF would like to thank Queen Mary University of London for support and the
Thomas Young Centre for a stimulating environment. AJM would like to thank
Dr Richard Wheatley and Dr J{\'a}nos {\'A}ngy{\'a}n for helpful comments and
Universit{\'e} de Lorraine for a visiting professorship during which part
of this work was completed.
We would like to thank Rory A. J. Gilmore for the SAPT(DFT) calculations on the methane dimer.

% Now include the BIBLIOGRAPHY.bib file that contains all references and uses
% the abbreviations from the above file.
\setlength\bibsep{2pt}
%\bibliography{macros,BIBLIOGRAPHY}
\newcommand{\SortNoOp}[1]{}

\end{document}